\documentclass[prd,superscriptaddress,amsfonts,amssymb,twocolumn,amsmath,showpacs]{revtex4-2}
\usepackage{bm}
\usepackage{amsfonts}
\usepackage{latexsym}
\usepackage[utf8]{inputenc}
\usepackage{graphicx}
\usepackage{amsmath}
\usepackage{palatino}
\usepackage{mathpazo}
\usepackage[british]{babel}
\usepackage{hhline}
\usepackage{multirow}
\usepackage{textcomp}
\usepackage{float}
\usepackage{booktabs}
\usepackage{dcolumn}
\usepackage{lipsum} 
\usepackage{mdframed}
\usepackage[]{mdframed}
\usepackage{hhline}
\usepackage{multirow}
\usepackage{ragged2e}
\usepackage{hyperref}
\hypersetup{colorlinks,citecolor=blue}
\hypersetup{colorlinks=true,linkcolor=red,filecolor=magenta,    urlcolor=cyan}
\usepackage{amsmath}
\usepackage{xcolor}
\usepackage{orcidlink}
\usepackage{epsfig}
\usepackage{caption}
\usepackage{subcaption}
\usepackage{commath}

\def\jnl@style{\it}
\def\aaref@jnl#1{{\jnl@style#1}}

\def\aaref@jnl#1{{\jnl@style#1}}

\def\aj{\aaref@jnl{AJ}}                   
\def\apj{\aaref@jnl{ApJ}}                 
\def\apjl{\aaref@jnl{ApJ}}                
\def\apjs{\aaref@jnl{ApJS}}               
\def\apss{\aaref@jnl{Ap\&SS}}             
\def\aap{\aaref@jnl{A\&A}}                
\def\aapr{\aaref@jnl{A\&A~Rev.}}          
\def\aaps{\aaref@jnl{A\&AS}}              
\def\mnras{\aaref@jnl{Mon.~Not.~Roy.~Astron.~Soc.}}             
\def\prd{\aaref@jnl{Phys.~Rev.~D}}        
\def\prc{\aaref@jnl{Phys.~Rev.~C}}  
\def\prl{\aaref@jnl{Phys.~Rev.~Lett.}}    
\def\qjras{\aaref@jnl{QJRAS}}             
\def\skytel{\aaref@jnl{S\&T}}             
\def\ssr{\aaref@jnl{Space~Sci.~Rev.}}     
\def\zap{\aaref@jnl{ZAp}}                 
\def\nat{\aaref@jnl{Nature}}              
\def\aplett{\aaref@jnl{Astrophys.~Lett.}} 
\def\apspr{\aaref@jnl{Astrophys.~Space~Phys.~Res.}} 
\def\physrep{\aaref@jnl{Phys.~Rep.}}      
\def\physscr{\aaref@jnl{Phys.~Scr}}       
\def\commat{\aaref@jnl{Comm.~Math.~Phys.}}              
\def\science{\aaref@jnl{Science}}               
\def\cqg{\aaref@jnl{Classical Quant.~Grav.}}            
\def\jpcs{\aaref@jnl{JPCS}}                                     
\def\ijmpd{\aaref@jnl{Int.~J.~Mod.~Phys.~D}}                    
\def\grg{\aaref@jnl{Gen.~Relat.~Gravit.}}               
\def\rpp{\aaref@jnl{Rep.~Prog.~Phys.}}          
\def\npa{\aaref@jnl{Nucl.~Phys.~A}}        
\def\lrr{\aaref@jnl{Living Rev.~Rel.}}                   
\def\jcap{\aaref@jnl{J.~Cosmology Astropart.~Phys.}}    
\def\rmp{\aaref@jnl{Rev.~Mod.~Phys.}}   
\def\epjc{\aaref@jnl{Eur.~Phys.~J.~C}} 
\def\plb{\aaref@jnl{~Phy.~Lett.~B}} 
\def\mpla{\aaref@jnl{Mod.~Phy.~Lett.~A}} 
\def\arxiv{\aaref@jnl{arxiv.org}}

\allowdisplaybreaks[1]

\begin{document}

\title{\bf Dynamical system analysis in teleparallel gravity with boundary term}

\author{S. A. Kadam\orcidlink{0000-0002-2799-7870}}
\email{k.siddheshwar47@gmail.com}
\affiliation{Department of Mathematics, Birla Institute of Technology and Science-Pilani, Hyderabad Campus, Hyderabad-500078, India}

\author{Ninaad P. Thakkar\orcidlink{0009-0003-3760-6436}}
\email{tninaad@gmail.com}
\affiliation{Department of Mathematics, Birla Institute of Technology and Science-Pilani, Hyderabad Campus, Hyderabad-500078, India}

\author{B. Mishra\orcidlink{0000-0001-5527-3565}}
\email{bivu@hyderabad.bits-pilani.ac.in}
\affiliation{Department of Mathematics, Birla Institute of Technology and Science-Pilani, Hyderabad Campus, Hyderabad-500078, India}


\begin{abstract} {\textbf{Abstract:}}
In this paper, we perform the dynamical system analysis of the cosmological models framed in the extended teleparallel gravity, the $f (T, B)$ gravity. We use the mapping, $f(T, B)\rightarrow-T+\Tilde{f}(T, B)$, and define the dynamical variables to form the autonomous dynamical system. The critical points are obtained in two well-motivated forms of $f (T, B)$, one that involves the logarithmic form of the boundary term $B$, and the other one is the non-linear form of the boundary term. The position of critical points is shown in the different evolutionary phases of the Universe such as radiation, matter, and de-Sitter phase. The stability condition of each of the critical points of both the models is derived and the behavior of each point has been obtained mathematically and through the phase portrait.
The evolution of standard density parameters such as radiation ($\Omega_{r}$), matter ($\Omega_{m}$), and dark energy ($\Omega_{DE}$) are also analyzed. Further to connect with the present cosmological scenario, the behavior of deceleration and equation of state parameter both in the dark energy phase ($\omega_{DE}$) and total ($\omega_{tot}$) are
shown from the initial condition of the dynamical variables. The accelerating behaviour has been obtained for both models.
\end{abstract}

\maketitle

\textbf{Keywords}: Teleparallel gravity, Dynamical system analysis, Boundary term, Phase portrait.

\section{Introduction}\label{Introduction}
General Relativity (GR) has been validated by decades of experimentation, and these experiments range from millimeter scale to solar system tests. These experiments are consistent with the emission of binary pulsars in terms of gravitational waves. It is assumed that in the standard model of cosmology, GR describes gravity at all scales. From type Ia supernovae observations \cite{Riess:1998cb, Perlmutter:1998np}, one can find that Universe is expanding faster than it used to be, which has been further supported by cosmic microwave background \cite{kogut2003} and the large-scale structure \cite{tegmark2004}. Though there are several candidates have been proposed, dark energy (DE) and dark matter provide some strong reasons. Hence, the effect of repulsive gravity has been introduced to explain the DE phenomena. It has less theoretical background and contrasts with the usual attractive nature of gravity. Therefore, in GR, the cosmological constant \cite{martin2012everything} has been introduced as an additional component. But when the cosmological terms are interpreted as a vacuum desired value, then it is encountered with the problems like fine-tuning and coincidence \cite{carroll2001}. So, to address this, the geometrical modification to GR has been proposed, which may explain the unknown nature of DE. 

One of the possible geometrical modifications to GR is Teleparallel Gravity (TG), which was initially proposed by Einstein as an alternative to GR \cite{unzicker2005translation}. The Lagrangian in TG consists of the torsion scalar $T$ term obtained through the contractions of the torsion tensor. The variation of this Lagrangian with respect to the tetrad gives rise to the evolution equations, which is the same as that of GR \cite{Clifton:2011jh, bahamonde:2021teleparallel}. So, TG is also known as the Teleparallel Equivalent of General Relativity (TEGR). One of the key factors that distinguish between GR and TEGR is the existence of tetrad fields. The tetrad fields can be useful to establish a linear Weitzenböck connection \cite{Weitzenbock1923, bahamonde:2021teleparallel}, which is a kind of connection related to torsion in the absence of curvature. Whereas, the curvature is used to geometrize space-time and represent gravitational interaction in GR. To be specific, in TEGR the gravitational interaction is defined by torsion \cite{Aldrovandi:2013wh,Ferraro:2008ey,bahamonde:2021teleparallel}. In recent times, TG and its extension have gained significant attention because of its ability to address some of the issues of the present Universe.

The TEGR Lagrangian contains the torsion scalar $T$, which can further be generalized to $f(T)$ theory \cite{Ferraro:2006jd, Cai:2015emx,Bengochea:2008gz,Duchaniya:2022rqu}. In $f(T)$ gravity, the field equations are of second order whereas $f(R)$ gravity is of fourth order \cite{Sotiriou:2008rp, Faraoni:2008mf, Capozziello:2011et}. In Ref. \cite{li2011f} it has been demonstrated that $f(T)$ gravity theory and its field equations are not invariant under local Lorentz transformations. So, $f(T)$ gravity can be generalised to $f(T, B)$ gravity \cite{Bahamonde:2015zma}, where $B$ is the boundary term. The $f(T,B)$ gravity has been studied in bouncing cosmology \cite{caruana2020}, in thermodynamical aspects  \cite{BAHAMONDE2018THERMODYNAMICS,pourbagher2019THERMODYNAMICS}, and in the cosmological evolution\cite{paliathanasis2021epjp}. We discuss some of the recent work pertaining to cosmological models in $f(T, B)$ gravity. In Noether symmetry, the cosmological model of $f(T, B)$ gravity has been analyzed in Ref. \cite{Bahamonde:2016grb}. The accelerating behaviour of the cosmological model has been shown in Refs. \cite{kadam2022accelerating1,kadam2022accelerating2}. Using cosmological observations, the cosmic expansion phenomena have been shown in Refs \cite{Franco:2021,briffa202}. Apart from this modification, the addition of Gauss-Bonnet invariant \cite{Kofinas:2014owa,Kofinas:2014aka,LOHAKARE2023101164,Kadam:2023}, the trace of energy-momentum tensor \cite{Harko2014,Bahamonde:2019shr}  and scalar field \cite{Gonzalez2020jss,duchaniya2023dynamical,duchaniya2023noetherscalartensor} in $f(T)$ gravity have shown promising results on the accelerated expansion of the Universe.

The objective of this study is to find out the stable critical points through dynamical system analysis and their corresponding cosmological behaviour in some well-defined form of $f(T, B)$.   The dynamical system analysis has been effective to study the asymptotic behavior and overall cosmic dynamics of the cosmological models. To note from the phase space and stability analysis, one can bypass the non-linearities of the cosmological equations \cite{dutta:2019,Narawade:2022aa,Agrawal:2023}. Also, by connecting the critical points to relevant evolutionary epochs, the description of global dynamics can be obtained. The phase of accelerated expansion at the late time generally corresponds to late time attractor, whereas the phase of radiation and matter dominance corresponds to saddle points \cite{duchaniya2023dynamical,Kadam:2022lgq}. Some of the relevant studies in $f(T,B)$ gravity can be seen in Refs. \cite{Franco:2021,Franco:2020lxx,samaddar2023EPJC}. Though several dynamical analysis studies are made in $f(T)$ gravity\cite{zhang2011notes,Mirza_2017}, we are motivated to analyse the study with the addition of boundary term $B$ i.e. in $f(T)$ to $f(T, B)$ gravity formalism. The paper is organised as: the background of $f(T, B)$ gravity and its field equations are presented in Sec.--\ref{sec:background}. In Sec.--\ref{sec:dynamicalsystemanalysis} the dynamical system analysis in $f(T, B)$ gravity has been presented with two forms of $f(T, B)$ leading to two models. The results and conclusions are given in Sec.--\ref{sec:conclusion}.

\section{\texorpdfstring{$f(T,B)$}{} Gravity Field Equations}\label{sec:background}

The teleparallel theories of gravity can be formed through the tetrad $e^{a}_{\mu}$ and its inverse, $ e_{a}^{\mu}$. It replaces the metric as the fundamental variable through the expressions, 
\begin{align}\label{metric_tetrad_rel}
    g_{\mu\nu}=e^{a}_{\ \ \mu} e^{b}_{\ \ \nu}\eta_{ab}\,,& &\eta_{ab} =  e_{a}^{\ \ \mu} e_{b}^{\ \ \nu}g_{\mu\nu}\,.
\end{align}
The Latin indices indicate the coordinates on the tangent space and the Greek indices represent the indices on the general manifold, which connect both the spaces and play a vital role in raising and lowering indices between the different spaces \cite{Cai:2015emx}. The tetrads with the metric adhere to the orthogonality conditions as, 
\begin{align}
    e^{a}_{\ \ \mu} e_{b}^{\ \ \mu}=\delta^a_b\,,&  & e^{a}_{\ \ \mu} e_{a}^{\ \ \nu}=\delta^{\nu}_{\mu}\,.
\end{align}
The flat spin connection  $ \omega^{a}_{\ \ b\mu}$  plays a role in including the local Lorentz transformation invariance in the equations of motion, which results from the appearance of the tangent space indices. Through the tetrad and spin connection, the teleparallel connection can be defined as\cite{Weitzenbock1923,Krssak:2015oua}
\begin{equation}
     \Gamma^{\sigma}_{\ \ \nu\mu} :=  e_{a}^{\ \ \sigma}\left(\partial_{\mu} e^{a}_{\ \ \nu} +  \omega^{a}_{\ \ b\mu} e^{b}_{\ \ \nu}\right)\,,
\end{equation}
Together, the tetrad as well as spin connection correspond to the gravitational and local degrees of freedom of the system and preserve the equations of motion's diffeomorphism and Lorentz invariance. Now, the torsion tensor can be derived from the teleparallel connection as \cite{Hayashi:1979qx} 
\begin{equation}
    T^{\sigma}_{\ \ \mu\nu} :=2 \Gamma^{\sigma}_{\ \ [\nu\mu]}\,,
\end{equation}
where the square bracket represents an antisymmetric operator. A torsion scalar can be defined using certain contractions of the torsion tensor \cite{Krssak:2018ywd,Cai:2015emx,Aldrovandi:2013wh,bahamonde:2021teleparallel},
\begin{equation}\label{eq:torsion_scalar_def}
    T:=\frac{1}{4}T^{\alpha}_{\ \ \mu\nu}T_{\alpha}^{\ \ \mu\nu} + \frac{1}{2}T^{\alpha}_{\ \ \mu\nu}T^{\nu\mu}_{\ \ \  \  \alpha} - T^{\alpha}_{\ \ \mu\alpha}T^{\beta\mu}_{\ \ \ \ \beta}\,,
\end{equation}
Torsion scalars are fundamentally dependent only on the teleparallel connection, as curvature scalars are dependent only on Levi-Civita connections. The Ricci scalar vanishes when calculated using the teleparallel connection, i.e., $R \equiv 0$. From this background one can write an expression that connects the regular Ricci scalar, $\mathring{R}=\mathring{R}(\mathring{\Gamma}^{\sigma}_{\ \ \mu\nu})$ (over-circles are used throughout to denote
quantities determined using the Levi-Civita connection) and torsion scalars $T$ as \cite{Bahamonde:2015zma,Farrugia:2016qqe}
\begin{equation}\label{LC_TG_conn}
    R=\mathring{R} + T - B = 0\,.
\end{equation}
where $B$ represents a total divergence term and is defined as
\begin{equation}\label{eq:boundary_term_def}
    B = \frac{2}{e}\partial_{\rho}\left(e T^{\mu \ \ \rho}_{\ \ \mu}\right)\,.
\end{equation}
The determinant of the tetrad, $e=\det\left(e^{a}_{\ \ \mu}\right)=\sqrt{-g}$.
Another extension of TEGR action can be written by incorporating an arbitrary function of both the torsion scalar $T$ and the boundary term $B$ \cite{Bahamonde:2015zma,bahamonde:2021teleparallel,Franco:2021},
\begin{equation}
S_{f(T,B)}=\int d^4x e\mathcal{L}_m+\frac{1}{2\kappa^2}\int d^4x  e f(T,B)\,,\label{actionequation}
\end{equation}
$\kappa^2= 8\pi G$. Varying action [ Eq.~\eqref{actionequation}] with respect to the tetrad fields, the $f(T,B)$ gravity field equations can be obtained as, 
\begin{eqnarray}
& & e_a{}^{\mu} \square f_B -  e_a {}^{\nu} \nabla ^{\mu} \nabla_{\nu} f_B +
	\frac{1}{2} B f_B e_a{}^{\mu} - \left(\partial _{\nu}f_B + \partial
	_{\nu}f_{T} \right)S_a{}^{\mu\nu}  \nonumber \\
& & -\frac{1}{e} f_{T}\partial _{\nu} (eS_a{}^{\mu\nu})
	+ f_{T} T^{B}{}_{\nu a}S_{b}{}^{\nu\mu}- f_T \omega ^b{}_{a\nu}
	S_b{}^{\nu\mu} -\frac{1}{2}  f E_{a}{}^{\mu} \nonumber\\ 
 & & = \kappa ^2  \Theta _a{}^{\mu} \,\label{eq:2}
\end{eqnarray}

The partial derivative of $f(T,B)$ with respect to $T$ and $B$ respectively denoted as $f_T$ and $f_B$; whereas the energy-momentum tensor and the Levi-Civita covariant derivative with respect to the Levi-Civita connection respectively denoted as, $\Theta_{a}^{~~\mu}$ and $\nabla_{\nu}$. We consider the tetrad for the flat FLRW metric as, 
\begin{equation}\label{eq:3}
e_{\ \ \mu}^{a}=(1,a(t),a(t),a(t))\,.
\end{equation}
This choice of tetrad allows for vanishing spin connection components,
$\omega^{a}_{\ \ b\mu}=0$, the above expression also satisfies the Weitzenb\"{o}ck gauge for $f(T,B)$ gravity. Here we shall study the cosmological aspects of  $f(T, B)$ gravity and hence consider the flat FLRW space-time as,  
\begin{equation}\label{eq:4}
ds^2=-dt^2+a^2(t)(dx^2+dy^2+dz^2)\,.
\end{equation}
Where $a(t)$ be the expansion along spatial directions. We can also find the Hubble parameter, $H\equiv\frac{\dot{a}}{a}$, with an over dot being the derivative of the variable in cosmic time $t$. In flat space-time, the torsion scalar $T$ and the boundary term $B$ respectively reduce to,
\begin{equation} \label{5}
T=6H^{2}, \hspace{2cm} B=6(3H^{2}+\dot{H}).
\end{equation}
Now, the field equations of $f(T,B)$ gravity \eqref{eq:2} for the metric \eqref{eq:4} and tetrad \eqref{eq:3} can be derived as, 
\begin{eqnarray}
   &-&3H^2(3f_B+2f_T)+3H \dot{f_B}-3\dot{H}f_B+\frac{1}{2}f(T,B)=\kappa^2\rho \nonumber\label{eq:6}\\
   &-&3H^2(3f_B+2f_T)-\dot{H}(3f_B+2f_T)-2H\dot{f_T} +\ddot{f_B}\nonumber\\
   &+&\frac{1}{2}f(T,B)=-\kappa^2 p\,. \label{eq:7}
\end{eqnarray}
One of the most important properties of this theory is, that it satisfies the continuity equation $(\dot{\rho}_i+3H(\rho_i+p_i)=0)$ for $i= m, r, DE$, that is matter, radiation, and the DE  respectively. To better understand the contributions of the modified Lagrangian, we consider the $f (T, B)$ gravity Lagrangian mapping, $f(T, B) \rightarrow - T + \Tilde{f}(T, B)$, then the Friedmann equations can be obtained as,\\
\begin{eqnarray}
    3H^{2}&=&\kappa^{2}\left(\rho+\rho_{DE}\right)\,,\\ \label{eq:10}
    3H^{2}+2\dot{H}&=&-\kappa^{2}\left(p+p_{DE}\right)\,.\label{eq:11}
\end{eqnarray} 
Where the expression for energy density and pressure for the DE phase can be obtained as,
\begin{eqnarray}
& &3H^2(3\Tilde{f}_B+2\Tilde{f}_T)-3H \dot{\Tilde{f}}_B+3\dot{H}\Tilde{f}_B\nonumber\\
& &-\frac{1}{2}\Tilde{f} (T,B)=\kappa^2\rho_{DE} \,, \label{eq:12}\\
& &-3H^2(3\Tilde{f}_B+2\Tilde{f}_T)-\dot{H}(3\Tilde{f}_B+2\Tilde{f}_T)-2H\dot{\Tilde{f}}_T+\ddot{\Tilde{f}}_B\nonumber\\
& & +\frac{1}{2}\Tilde{f}(T,B)=\kappa^2 p_{DE}\,.\label{eq:13}
\end{eqnarray}
The expression of the equation of state (EoS) parameter for the DE phase can be written as,

\begin{equation}
    \omega_{DE}=-1+\frac{\ddot{\Tilde{f}}_{B}-3H\dot{\Tilde{f}}_{B}-2\dot{H}\Tilde{f}_{T}-2H\dot{\Tilde{f}}_{T}}{3H^{2}(3\Tilde{f}_{B}+2\Tilde{f}_{T})-3H\dot{\Tilde{f}}_{B}+3\dot{H}\Tilde{f}_{B}-\frac{1}{2}\Tilde{f}(T,B)}\,.\label{eq:14}
\end{equation}
Next, we shall define the dynamical variables and express the cosmological parameters in terms of dynamical variables. The cosmological behaviour of the models will be studied through some functional forms of $\Tilde{f}(T, B)$. 

\section{Dynamical System Analysis}\label{sec:dynamicalsystemanalysis}
From the theoretical point of view, any proposed cosmological model should contain at least part that explains "Inflation $\rightarrow$ Radiation $\rightarrow$ Matter $\rightarrow$ DE " \cite{bohmer2017dynamicalsystem}. To attain the aforementioned suggested cosmological model, inflation must be an unstable point in order for the Universe to have an inflation exit, whereas radiation and matter points must be saddle points in order for these eras to be long enough. The final phase of the DE era should be a stable period of accelerated expansion.
In order to analyse this, we consider the Universe filled with two fluids such that, $\rho=\rho_{m}+\rho_{r}$, where $\rho_m$ and $\rho_{r}$ respectively be the energy density for matter and radiation. In the matter dominated phase $p_{m}=0$ and hence $\omega_m$ vanishes, whereas in the radiation phase, $\omega_{r}=\frac{1}{3}$. With these, we define the dynamical variables as,
\begin{eqnarray}
& & X=\tilde{f}_{B} \,,\quad Y=\tilde{f}_{B} \frac{\dot{H}}{H^2}\,,\quad Z=\frac{\dot{\tilde{f}}_B}{H} \,,\quad 
V=\frac{\kappa^2 \rho_r}{3H^2}\,,\quad \nonumber\\
& & W=-\frac{\tilde{f}}{6H^2}\,.\label{dynamical variables}
\end{eqnarray}
The standard density parameters expressions for matter $(\Omega_m)$, radiation $(\Omega_r)$ and DE $(\Omega_{DE})$ phase are respectively,
\begin{align}
\Omega_{m}=\frac{\kappa^2 \rho_{m}}{3H^2}, \quad \Omega_{r}=\frac{\kappa^2 \rho_{r}}{3H^2}, \quad \Omega_{DE}=\frac{\kappa^2 \rho_{DE}}{3H^2}.
\end{align}
with
\begin{align}
\Omega_{m}+\Omega_{r}+\Omega_{DE}=1
\end{align}
So, the constrained equation in terms of a dynamical variable can be written as,
\begin{align}
\Omega_{m}+\Omega_{r}+W+2\tilde{f}_{T}+Y+3X-Z=1
\end{align}
where
\begin{align}
    \Omega_{DE}=W+2\tilde{f}_{T}+Y+3X-Z
\end{align}
The total EoS parameter and DE EoS parameter are respectively obtained in dynamical variables as,
\begin{align}
\omega_{tot}&=-1-\frac{2 Y}{3 X}\,,\nonumber\\
\omega_{DE}&=-\frac{(V+3) X+2 Y}{3 X (2 \tilde{f}_{T} +W+3 X+Y-Z)}\,.
    \label{expressionomegadeomegatot}
\end{align}
Subsequently, we have obtained the autonomous dynamical system as follows,
\begin{eqnarray}
X^{'}&=&Z\,,\nonumber\\
Y^{'}&=&\lambda  X+\frac{Y (Z-2 Y)}{X}\,,\nonumber\\
Z^{'}&=&6 \tilde{f}_{T} -V+3 W+\frac{2 \tilde{f}_{T}  Y}{X}-\frac{Y Z}{X}-\frac{2 Y}{X}\nonumber\\&+& 9 X+3 Y+2\tilde{f}^{\,'}_{T}-3\,,\nonumber\\
V^{'}&=&-\frac{2 V (2 X+Y)}{X}\,,\nonumber\\
W^{'}&=&-\frac{2 W Y}{X}-\lambda  X-\frac{2 \tilde{f}_{T}  Y}{X}-6 Y\,.\label{GDS}
\end{eqnarray}
Where $'$ denotes differentiation with respect to $N=ln(a)$, to express the autonomous dynamical system, we define the parameter $\lambda=\frac{\ddot{H}}{H^3}$ \cite{Odintsov:2018,Franco:2020lxx} and is treated as a constant throughout the analysis. To note, the value of the parameter $\lambda=8, \frac{9}{2},$ connects with the radiation, matter dominated phase respectively whereas for DE it depends on the dynamical variables $X$ and $Y$. Now, to study the stability analysis, we need some form of $\Tilde{f}(T, B)$, and hence we have considered two forms of $\Tilde{f}(T, B)$ that lead to two models.

\subsection{ Model-I} \label{ModelI}
We consider,
\begin{equation*}
\Tilde{f}(T,B) =\xi T + \alpha B log(B)    
\end{equation*}
This specific form of $f(T, B)$ has been successful in addressing the late time cosmic phenomena issue \cite{Escamilla-Rivera:2019ulu,bamba2011eos}, Noether symmetry \cite{Capozziello:2016eaz}. Also, the critical points can be analysed in the presence of a non-canonical scalar field and the exponential potential function in $f(T, B)$ gravity framework \cite{paliathanasis2021epjp}. In the absence of a scalar field, the cosmological aspects through the behavior of critical points analysis may provide some deeper insight into the evolution of the Universe in different evolution phases. The dynamical variable $Z$ from Eq.(\ref{dynamical variables}) can be written as, $Z=\alpha\left[\frac{6Y+X\lambda}{3X+Y}\right]$ and treated as the dependent variable. The autonomous dynamical system for this setup can be obtained as,
\begin{align}
X^{'}&=\frac{\alpha  (\lambda  X+6 Y)}{3 X+Y}\,,\nonumber\\
Y^{'}&=X \left(\lambda -\frac{2 Y^2}{X^2}\right)+\frac{\alpha  Y (\lambda  X+6 Y)}{X (3 X+Y)}\,,\nonumber\\
V^{'}&=-\frac{2 V (2 X+Y)}{X}\,,\nonumber\\
W^{'}&=-\frac{2 W Y+\lambda  X^2+6 X Y+2 \xi  Y}{X}\,.
    \label{Dynamicalsystemmodel-I}
\end{align}

The standard density parameters for DE and matter can be calculated as
\begin{align}
    \Omega_{DE}&=2 \xi +W+3 X+Y-\frac{\alpha  (\lambda  X+6 Y)}{3 X+Y}\nonumber\,,\\ 
    \Omega_{m}&=-2 \xi -V-W-3 X-Y+1+\frac{\alpha  (\lambda  X+6 Y)}{3 X+Y}\,.
\end{align}
and the EoS parameter for DE is given as.
\begin{align}
   \omega_{DE} =-\frac{(V+3) X+2 Y}{3 X \left(2 \xi +W-\frac{\alpha  (\lambda  X+6 Y)}{3 X+Y}+3 X+Y\right)}\nonumber\,.
\end{align}

 We shall find the critical points of the dynamical system by considering $X^{'}=0, Y^{'}=0, V^{'}=0, W^{'}=0$. The critical points and their existence conditions are given in Table \ref{modelIcriticalpoint},
 \begin{widetext}
\begin{table}[H]
\centering 
\begin{tabular}{|c|c|c|c|c|c|c|} 
\hline\hline 
 \parbox[c][0.8cm]{3.3cm}{\textbf{Name of Critical Point}} & $\textbf{X}$ & $\textbf{Y}$  & $\textbf{V}$ & $\textbf{W}$ & \textbf{Exist for}\\ [0.5ex] 
\hline\hline 
 \parbox[c][1cm]{3.6cm}{\textbf{$C_1= (X_1,Y_1,V_1,W_1)$}} & $X_1$ & $-2 X_1$ & $V_1$ &$-\xi-X_1$ & $X_1\neq 0,  \alpha=0, \lambda =8 $ \\
\hline
\parbox[c][1cm]{3.6cm}{$C_2= (X_2,Y_2,V_2,W_2)$} & $X_{2}$ & $-\frac{3}{2} X_2 $ & $0$  & $W_2$&  $X_2\neq 0, \xi=-W_2-3 X_2-Y_2, \alpha=0, \lambda =\frac{9}{2}.$\\
\hline
\parbox[c][1cm]{3.6cm}{$C_3= (X_3,Y_3,V_3,W_3)$} & $X_3$ & $Y_3$ & $0$  &$-3 X_3-Y_3-\xi $& $X_3\neq 0, Y_3 \left(3 X_3+2 Y_3\right)\neq 0, \alpha=0, \lambda =\frac{2 Y_3^2}{X_3^2}.$ \\
\hline
\parbox[c][1cm]{3.6cm}{$C_4= (X_4,Y_4,V_4,W_4)$} & $X_4$ & $0$ & $0$  &$W_4$& $X_4\neq 0,W_4 , \alpha, \xi=\text{arbitrary}, \lambda=0.$ \\
\hline
\end{tabular}
\caption{The critical points ({\bf Model-I})}
\label{modelIcriticalpoint}
\end{table}
\end{widetext}
The eigenvalues of the Jacobian matrix at each critical point have been obtained to analyse the stability of the critical points. The stability can be categorised as: (i) all the eigenvalues of the Jacobian matrix are negative, stable node; (ii) all the eigenvalues are positive, unstable node; (iii) the eigenvalues are both positive and negative, saddle node. In addition, at the stable spiral node, there is a negative determinant for the Jacobian matrix, and the real component of all of the eigenvalues also has a negative value. We have given the analysis of each of the critical points below:

\begin{table}[H]
    \centering 
    \begin{tabular}{|c|c|} 
    \hline\hline 
    \parbox[c][0.3cm]{1.3cm}{\textbf{C. P.}}& \textbf{Eigenvalues } \\ [0.5ex] 
    \hline\hline 
    \parbox[c][0.3cm]{1.3cm}{$C_1$ } & $\{0,0,4,8\}$   \\
    \hline
    \parbox[c][0.3cm]{1.3cm}{$C_2$ } & $\left\{6,3,-1,0\right\}$  \\
    \hline
   \parbox[c][0.3cm]{1.3cm}{$C_3$ } &  $\left\{0,-4\frac{Y_3}{X_3}, -2\frac{Y_3}{X_3}, -2\frac{(2X_3+Y_3)}{X_3}\right\}$ \\
   \hline
   \parbox[c][0.3cm]{1.3cm}{$C_4$ } &  $\{0,0,0,-4\}$   \\
    \hline
    \end{tabular}
    \caption{Eigenvalues corresponding to the critical point ({\bf Model-I})}
    \label{modelIeigenvalues}
\end{table}

\begin{figure}[H]
    \centering
    \includegraphics[width=75mm]{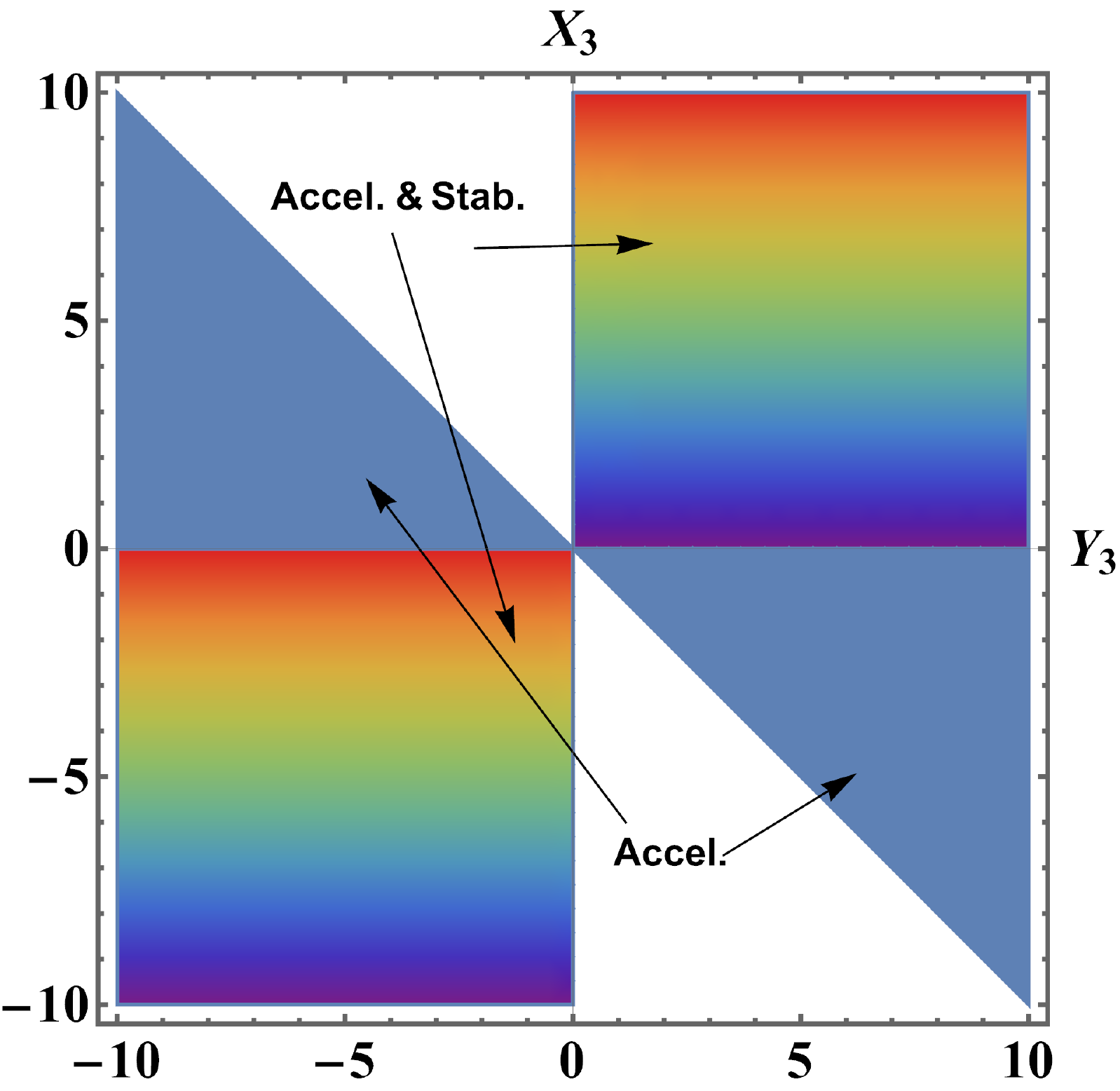}
    \caption{Stability and acceleration region of the critical point $C_{3}$ representing DE energy dominated era ({\bf Model-I}). } \label{modelIregionplot}
\end{figure}

\begin{figure}[H]
    \centering
    \includegraphics[width=58mm]{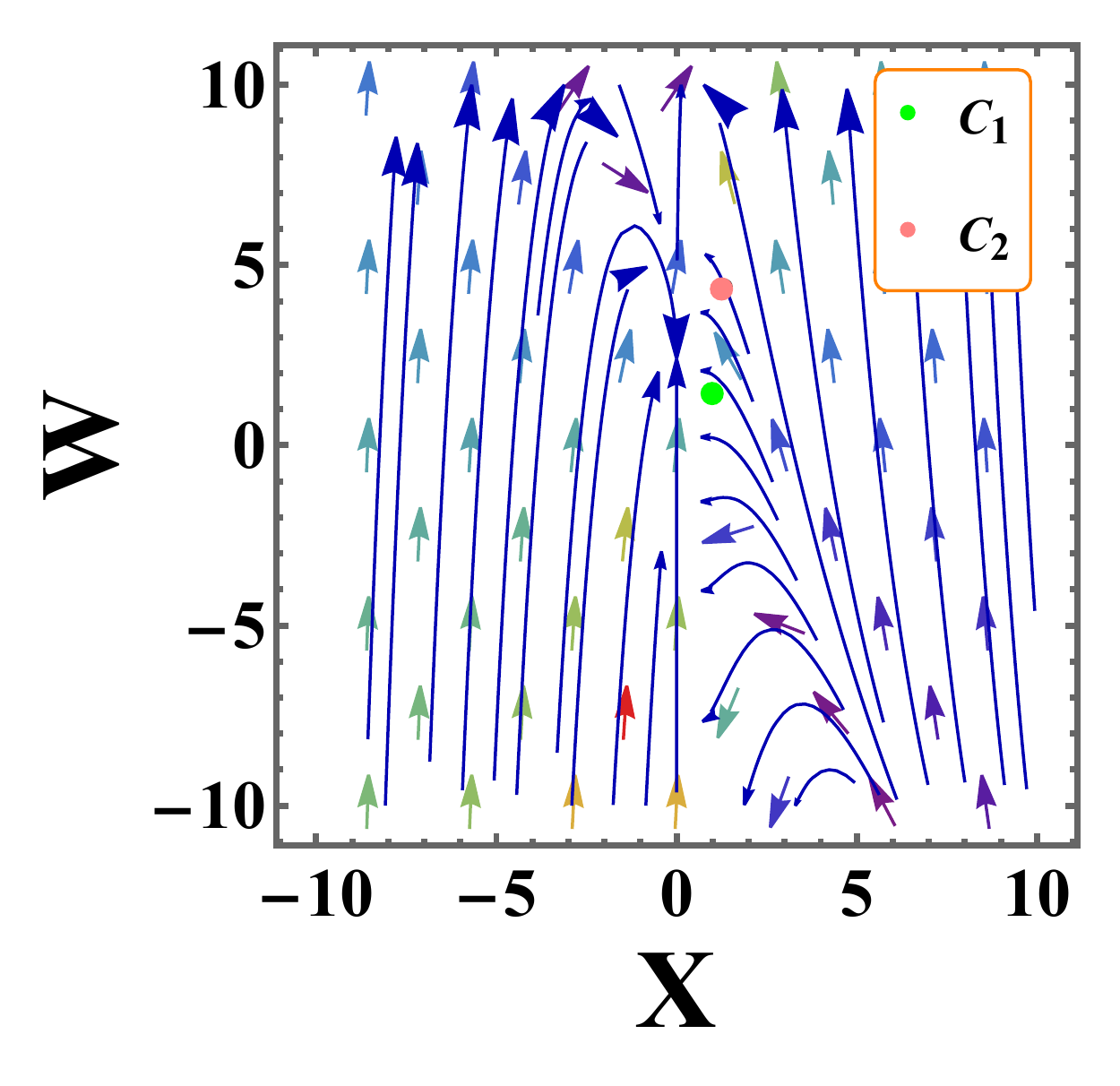}
    \includegraphics[width=58mm]{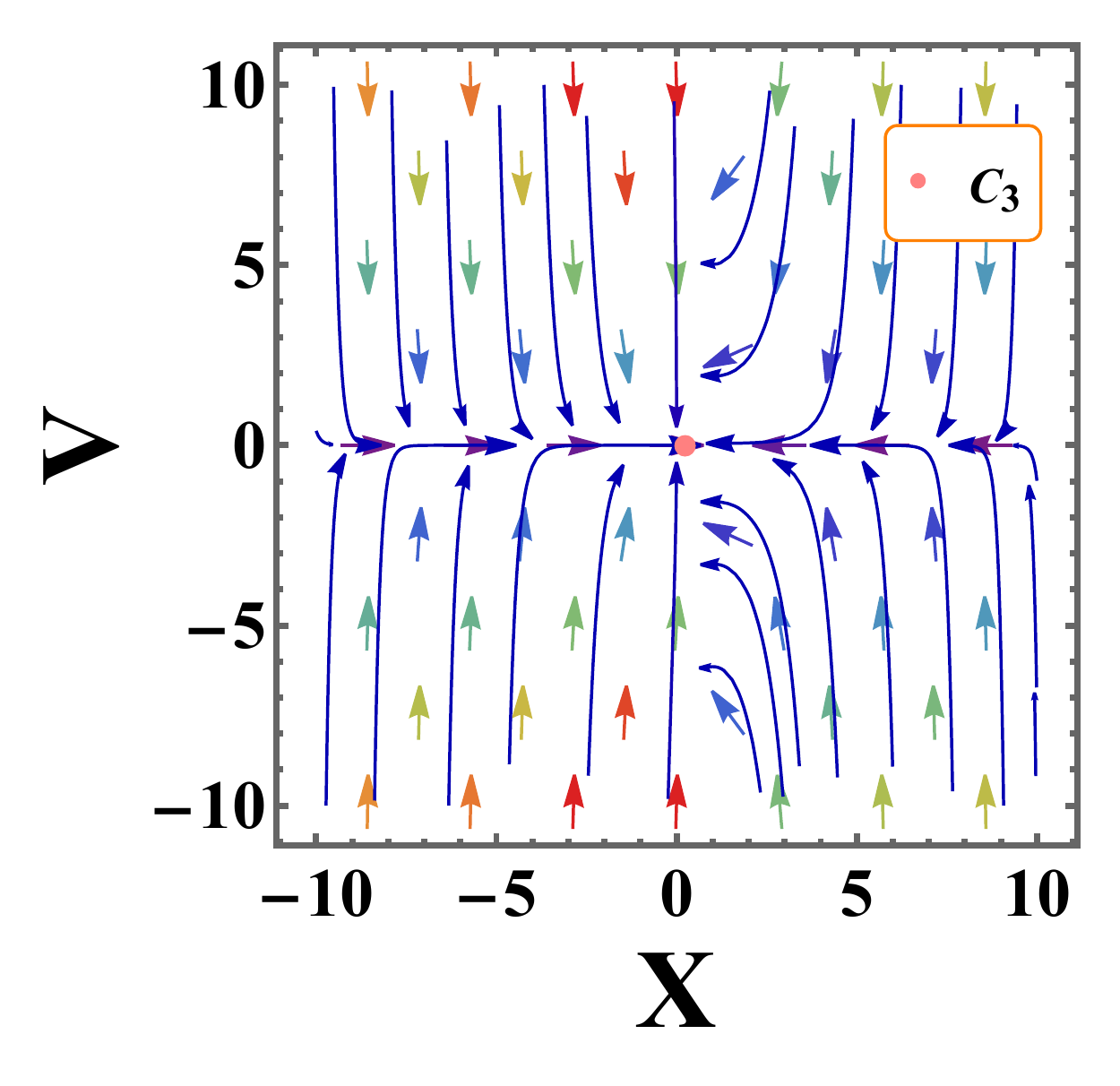}
    \caption{ $2D$ phase portrait for the dynamical system for $\lambda=0.3, \xi=-2.4, \alpha=1.3$ ({\bf Model-I}) .} \label{modelI2dphaseportrait}
\end{figure}

\begin{widetext}
\begin{center}
\begin{table}[H]
    \centering 
    \begin{tabular}{|c|c|c|c|c|} 
    \hline\hline 
    \parbox[c][0.8cm]{1.3cm}{ \textbf{C.P.}} & \textbf{Stability Conditions} & $\textbf{q}$ & \textbf{$\omega_{tot}$} & \textbf{$\omega_{DE}$} \\ [0.5ex] 
    \hline\hline 
    \parbox[c][0.8cm]{1.3cm}{\textbf{$C_1$}} & Unstable & $1$ & $\frac{1}{3}$ & $-\frac{V_1-1}{3 \xi }$ \\
    \hline
    \parbox[c][0.8cm]{1.3cm}{$C_2$}  &  Unstable& $\frac{1}{2}$ & $0$& 0\\
    \hline
    \parbox[c][0.8cm]{1.3cm}{$C_3$}  &  $\begin{tabular}{@{}c@{}}     Stable for\\$\left(X_3<0\land Y_3<0\right)\lor$\\$ \left(X_3>0\land Y_3>0\right)$\end{tabular}$ & $-\frac{X_3+Y_3}{X_3}$  & $-1-\frac{2 Y_3}{3 X_3}$ & $\frac{3 X_3+2 Y_3}{6 Y_3-3 \xi X_3}$ \\
     \hline
    \parbox[c][0.8cm]{1.3cm}{$C_4$}  &    Nonhyperbolic  & $-1$  & $-1$ & $-\frac{1}{2 \xi+W_4+3 X_4}$ \\
    [1ex] 
    \hline 
    \end{tabular}
     \caption{Stability condition, EoS parameter and deceleration parameter ({\bf Model-I})}
    \label{modelIstabilitycondi}
\end{table}
\end{center}
\end{widetext}
\begin{widetext}
\begin{center}
\begin{table}[H]
    \centering 
    \begin{tabular}{|c|c|c|c|c|c|} 
    \hline\hline 
   \parbox[c][0.6cm]{1cm}{\textbf{C. P.}} & \textbf{Evolution Eqs.} & \textbf{Universe phase} & $\Omega_{m}$& $\Omega_{r}$ & $\Omega_{DE}$\\ [0.5ex] 
    \hline\hline 
   \parbox[c][0.8cm]{1.3cm}{ $C_1$ }& $\dot{H}=-2 H^{2}$ & $a(t)= t_{0} (2 t+c_{2})^\frac{1}{2}$ & $-\xi-V_1+1$& $V_1$ & $\xi$\\
    \hline
    \parbox[c][0.8cm]{1.3cm}{$C_2$} & $\dot{H}=-\frac{3}{2}H^{2}$ & $a(t)= t_{0} (\frac{3}{2}t+c_{2})^\frac{2}{3}$ & $W_2+\frac{3 X_2}{2}+1$& $0$ & $-W_2-\frac{3 X_2}{2}$\\
    \hline
    \parbox[c][0.8cm]{1.3cm}{$C_3$ }&  $\dot{H}=-\frac{Y_3}{ X_3}H^2$ & $a(t)= t_{0} (\frac{Y_3}{X_3}t+c_{2})^\frac{X_3}{Y_3}$ & $-\xi +\frac{2 Y_3}{X_3}+1$& $0$ & $\xi -\frac{2 Y_3}{X_3}$\\
    \hline
    \parbox[c][0.8cm]{1.3cm}{$C_4$ }&  $\dot{H}=0$ & $a(t)= t_{0} e^{c_2 t}$ & $-2 \xi -W_4-3 X_4+1$& $0$ & $2 \xi +W_4+3 X_4$\\
    [1ex] 
    \hline 
    \end{tabular}
     \caption{Evolution Eqs., phase of the Universe, density parameters ({\bf Model-I})}
    \label{modelIdensityparameter}
\end{table}
\end{center}
\end{widetext}
\begin{itemize}
\item {Critical Point $C_1$:}
The critical point $C_1$ with $\lambda=8$ describes the radiation-dominated era. The value of the parameter $\omega_{tot}=\frac{1}{3}$, $q=1$. This critical point will describe the standard radiation-dominated era for $V_{1}=1,\xi=0$ at which the contribution of the standard density parameter for DE will vanish. The eigenvalues at this critical point are presented in Table \ref{modelIeigenvalues}, which shows this critical point is unstable in nature and the corresponding value of the deceleration and EoS parameter is that of radiation-dominated [Table \ref{modelIstabilitycondi}]. The evolution equation, along with the exact solution, is obtained and presented in Table \ref{modelIdensityparameter}. The exact solution obtained at this critical point is in the power law $a(t)=t_0 (t)^h$ form with $h=\frac{1}{2}$ which explains the radiation-dominated era. The behavior of phase space trajectories at this critical point shows that this critical point is the saddle point and hence unstable, as can be seen in Fig.  \ref{modelI2dphaseportrait}.

\item{Critical Point $C_2$:}
The value of the EoS parameter ($\omega_{tot}$) vanishes at this critical point, hence, this critical point represents the CDM (Cold Dark Matter) phase of the evolution of Universe. The value of dynamical variable $\lambda$ is $\frac{9}{2}.$ This critical point describes a non-standard CDM-dominated era with the small contribution of the DE density parameter [Refer Table \ref{modelIdensityparameter}]. The evolution equation with the exact solution at this critical point is presented in Table \ref{modelIdensityparameter}, the power law solution with index $\frac{2}{3}$ indicates that this will describe the CDM-dominated era. The existence of positive and negative eigenvalues at this critical point, as presented in Table \ref{modelIeigenvalues} shows that this critical point is unstable. The phase space trajectories at this critical point show that this critical point is a saddle point, which can be analyzed in Fig. \ref{modelI2dphaseportrait}, which will support the stability condition obtained from the sign of the eigenvalues.

\item {Critical Point $C_3:$} 
At this critical point, the value of $q, \omega_{DE}$ and $\omega_{tot}$ is dependent on the dynamical variables $X, Y$ hence, this critical point can describe the early as well as the late phases of the Universe evolution. The critical point will describe the de-sitter solution at $Y_3=0$ and describe accelerating expansion of the universe at parametric range $\left(X_3<0\land Y_3<-X_3\right)\lor \left(X_3>0\land Y_3>-X_3\right)$. To get better clarity, the range of parameters where it describes accelerating expansion and stability is plotted in the region plot in Fig. \ref{modelIregionplot}. From this, we can analyze that the value of $Y_3$ is near 0 in stability, existence, and in the parametric range where parameters are capable to describe the accelerated expansion of the Universe at critical point $C_3$, hence this critical point is capable of describing a DE dominated era of Universe evolution. The phase space trajectories at this critical point are attractors and can be analyzed in Fig. \ref{modelI2dphaseportrait}. This critical point is describing the standard DE-dominated era with $\Omega_{DE}=1$ at $\xi=1, Y_3=0$ (Refer Table \ref{modelIdensityparameter}). The exact solution obtained at this critical point is in the power law $a(t)=t_0 (t)^h$ form with $h=\frac{X_3}{Y_3}$ depending upon the value of dynamical variables $X_3$ and $Y_3$, the corresponding phase of the Universe evolution can be analysed. The eigenvalues at this critical point are normally hyperbolic \cite{coley2003dynamical} and are stable in the parameter range as described in Table \ref{modelIstabilitycondi}.
\end{itemize}
\begin{figure}[H]
    \centering
    \includegraphics[width=58mm]{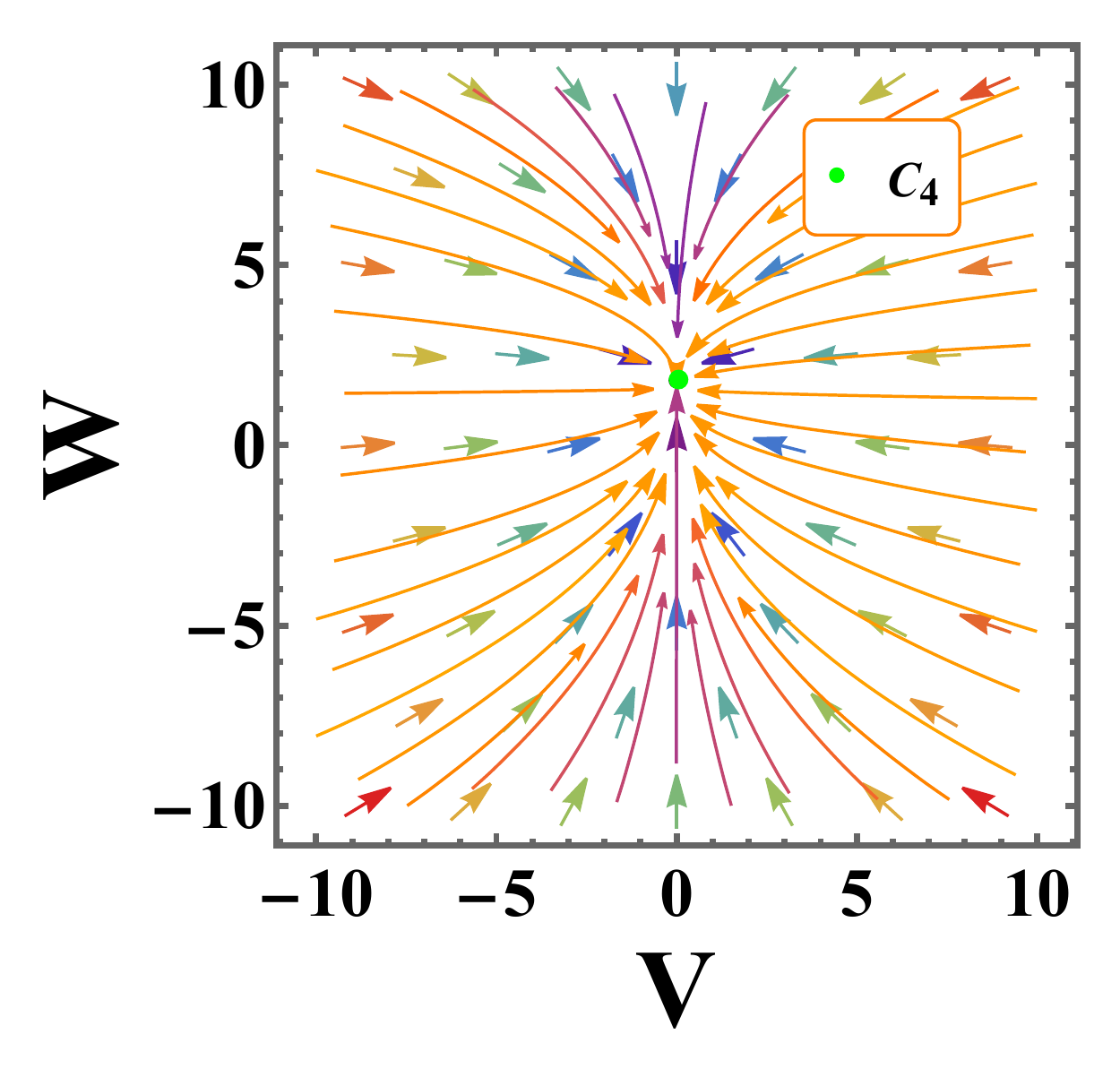}
    \caption{ $2D$ phase portrait for the dynamical system for $\lambda=0.3, \xi=-2.4, \alpha=1.3$ ({\bf Model-I}) .} \label{modelI2dphaseportrait2}
\end{figure}

\begin{figure}[H]
    \centering
    \includegraphics[width=83mm]{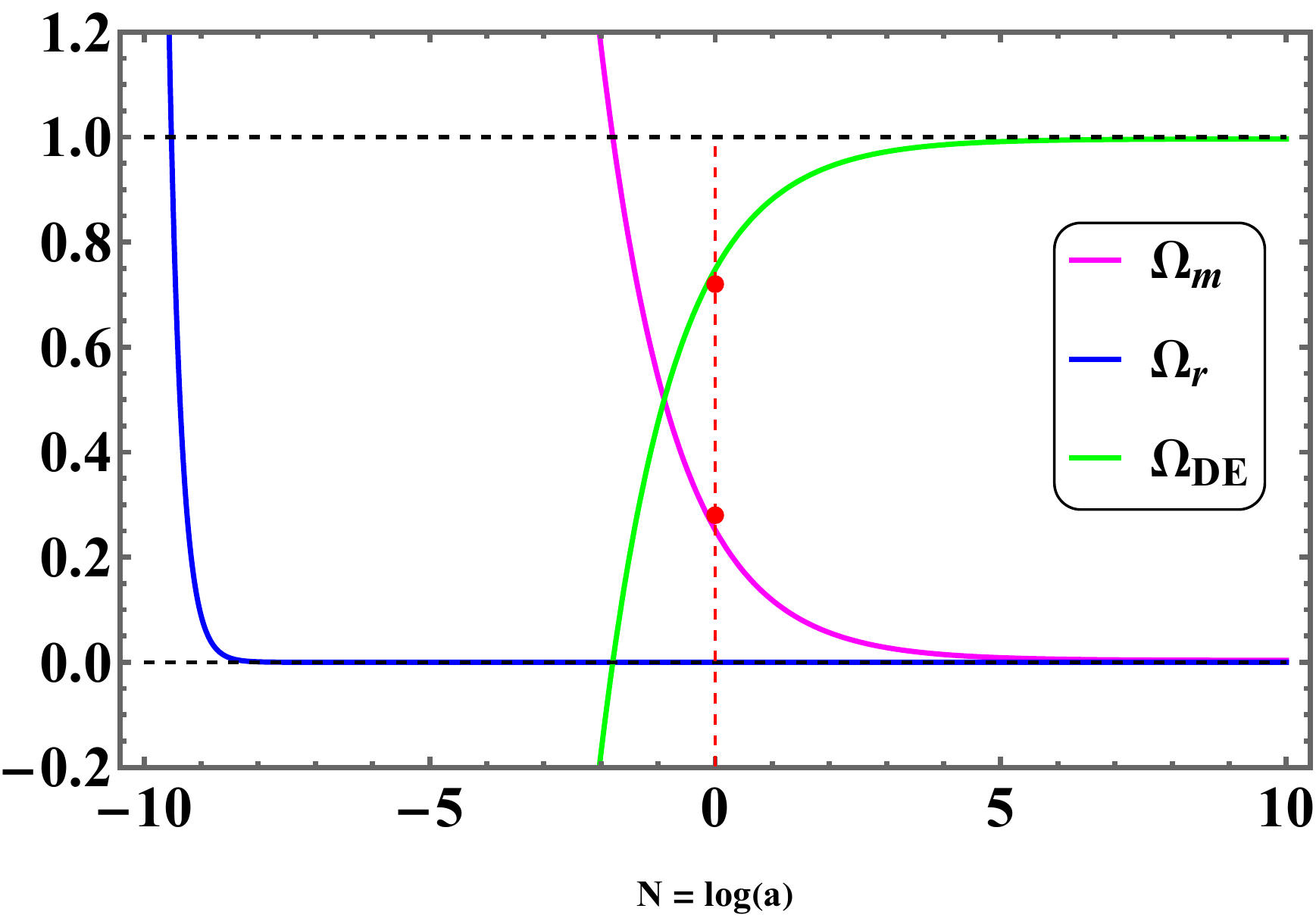}
    \caption{Evolution of density parameters. The initial conditions: $X=-1.2 \times 10^{2.2},\, Y=2.2 \times 10^{-3.4},\,  V=1.02 \times 10^{2.6},\, W=4.5\times 10^{-8.1},\,  \lambda=0.3,\, \xi=-2.4, \alpha=1.3$ ({\bf Model-I}).} \label{model1evolution}
\end{figure}

\begin{figure}[H]
    \centering
    \includegraphics[width=83mm]{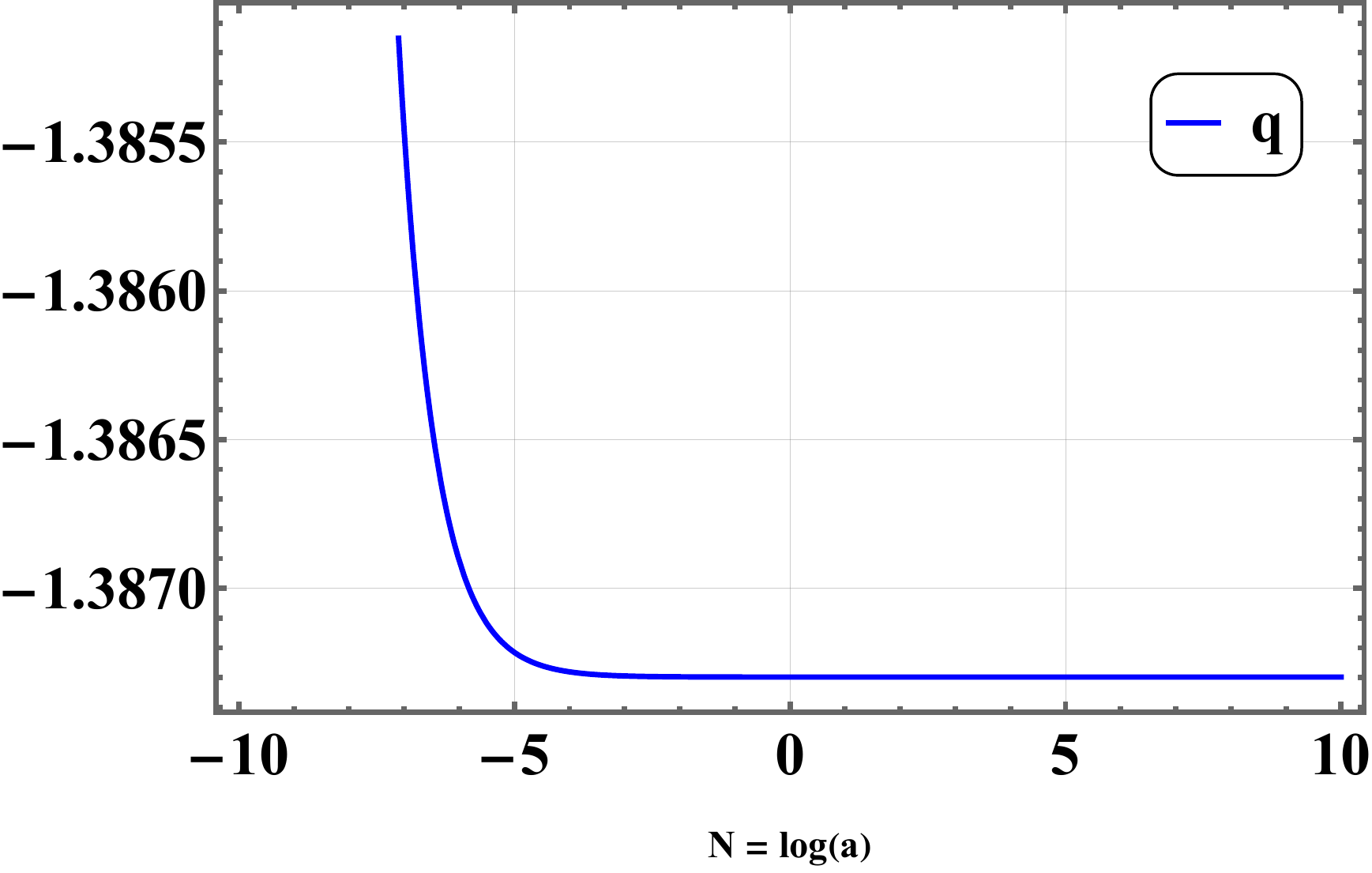}
    \caption{Deceleration parameter $(q)$ for the initial conditions $X=-1.2 \times 10^{2.2},\, Y=2.2 \times 10^{-3.4},\,  V=1.02 \times 10^{2.6},\, W=4.5\times 10^{-8.1},\,  \lambda=0.3,\, \xi=-2.4, \alpha=1.3$ ({\bf Model-I}).} \label{modelIomegadedecele}
\end{figure}

\begin{figure}[H]
    \centering
    \includegraphics[width=75mm]{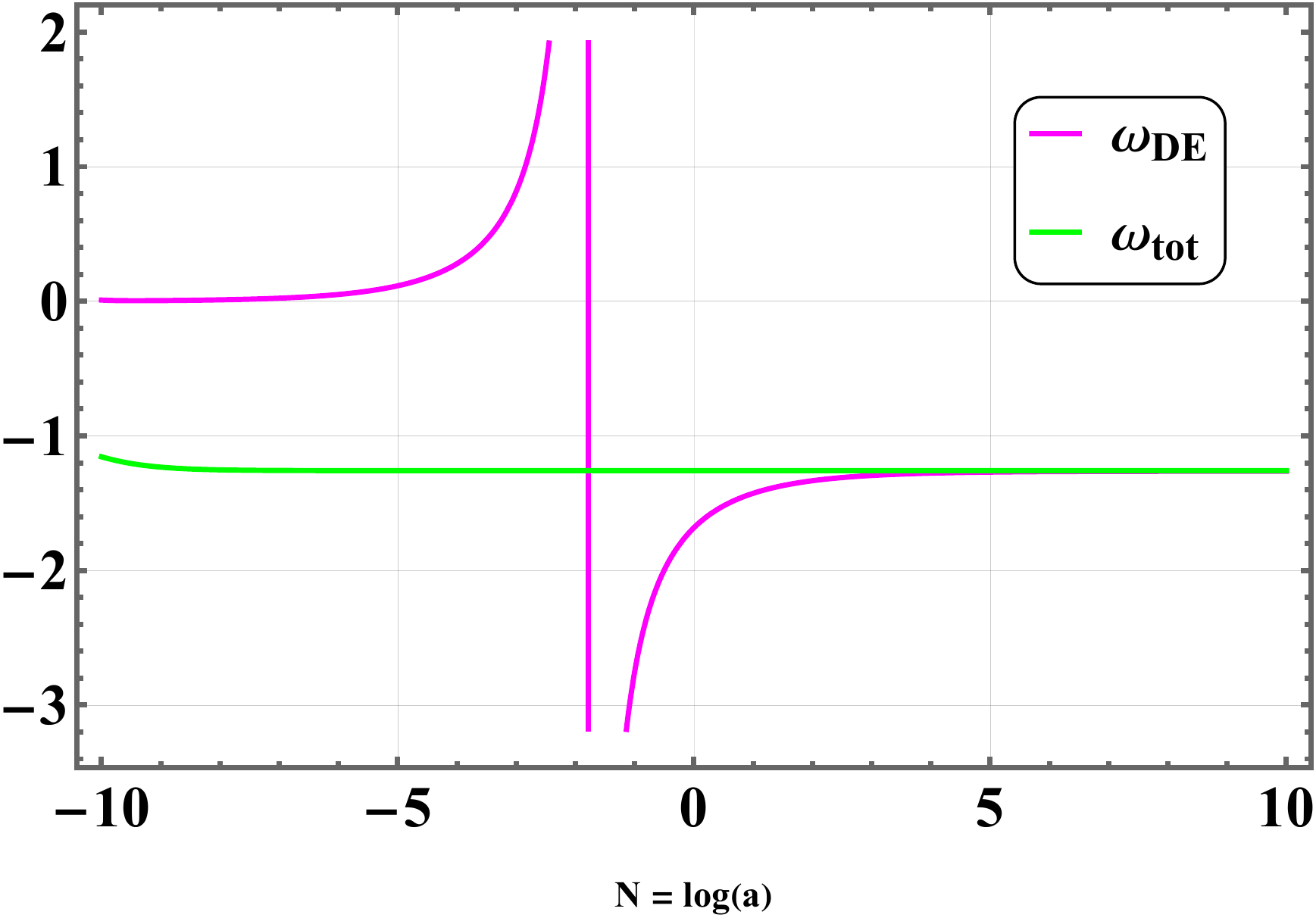}
    \caption{EoS parameters $\omega_{DE}$, $\omega_{tot}$ for the initial conditions $X=-1.2 \times 10^{2.2},\, Y=2.2 \times 10^{-3.4},\,  V=1.02 \times 10^{2.6},\, W=4.5\times 10^{-8.1},\,  \lambda=0.3,\, \xi=-2.4, \alpha=1.3$ ({\bf Model-I}).} \label{modelIomegadedecele2}
\end{figure}

\begin{itemize}
     \item {Critical Point $C_4:$} The value of $q=-1$, $\omega_{tot}=-1$ at this critical point, hence this critical point explains the de-Sitter solution. The exact cosmological solution at this critical point is described in Table \ref{modelIdensityparameter} which takes the de-Sitter solution form. From this one can observe that this critical point should explain a standard DE-dominated era at $\xi=\frac{1}{2}, W_4 =0, X_4 =0$. But due to the presence of zero eigenvalues, this critical point is non-hyperbolic in nature. The linear stability theory fails to provide information regarding the stability of the critical point if the critical point contains zero eigenvalues. Moreover, since the system equations are not satisfying the central manifold condition (after separating from linear and non-linear parts, the non-linear part at zero is not vanishing), therefore it fails to describe the stability of this critical point. The 2-D phase space diagram to analyse the behaviour of the phase space trajectories at this critical point in Fig. \ref{modelI2dphaseportrait2} has been given, where it is observed that the phase space trajectories are attracted towards this critical point. Hence this critical point is an attractor.
\end{itemize}
In Fig. \ref{modelIregionplot}, the stability and acceleration region of the critical point $C_3$ has been shown. Though the light blue shaded area in the figure shows the acceleration, the stability could not be established. The evolution plot of standard density parameters in terms of the $N=log(a)$ has been given in Fig. \ref{model1evolution}, we can find the value of $\Omega_{r} \approx 0.019, \Omega_{m} \approx 0.3$ and $\Omega_{DE} \approx 0.7$. The vertical dashed line represents the present time. The blue curve represents the evolution of the standard density parameter for radiation, and it can be observed that this curve dominates the other two curves at the early evolution and decreases gradually from early to late time and tend to zero at late times. The behavior of $q$ and Eos for $\omega_{tot}, \omega_{DE}$ can be analyzed in Figs. \ref{modelIomegadedecele} and \ref{modelIomegadedecele2} respectively. Since the plot of $q$ lies in the negative region, hence it is capable to describe the current accelerated expansion of the Universe. The value of $q$ at the present time is $-1.387$ which is approximately the same as the current observation study \cite{Feeney:2018}. The plot of $\omega_{tot}$ at present value takes the value $-1.233$ which agrees with $\omega_0=-1.29 ^{+0.15}_{-0.12}$\cite{DIVALENTINO:2016}.



\subsection{ Model-II} \label{ModelII}
We consider,
\begin{equation*}
\Tilde{f}(T,B) = \zeta T + \beta (-B)^p 
\end{equation*}
where $\Tilde{f_2}(B)= \beta (-B)^p$ in specifically, is nonlinear and capable of studying observational tests for the theory by referring to the most recent SN-Ia data \cite{Bengochea:2008gz}. This is a prominent form of $\tilde{f}(T, B)$ gravity due to its usefulness in explaining the present cosmic expansion and $H_0$ tension\cite{briffa202}. Similar to the first model, we get a relationship for the dynamical variable $Z=\left[\frac{(p-1) X (\lambda  X+6 Y)}{3 X+Y}\right],$ hence the dynamical variable $Z$ is treated as a dependent and the others are independent. The terms $\tilde{f}_{T}=\zeta $ and $\tilde{f}^{\,'}_{T}=0\,$ will convert the general dynamical system in Eq. (\ref{GDS}) into an autonomous form as follows,

 \begin{align}
X^{'}&=\frac{(p-1) X (\lambda  X+6 Y)}{3 X+Y}\,,\nonumber\\
Y^{'}&=\frac{(p-1) Y (\lambda  X+6 Y)}{3 X+Y}+X \left(\lambda -\frac{2 Y^2}{X^2}\right)\,,\nonumber\\
V^{'}&=-\frac{2 V (2 X+Y)}{X}\nonumber\\
W^{'}&=-\frac{2 W Y+\lambda  X^2+6 X Y+2 \zeta  Y}{X}\,.
    \label{Dynamicalsystemmodel-II}
\end{align}
The standard density parameter for DE and matter can be given as,
\begin{align}
    \Omega_{DE}&=2 \zeta -\frac{(p-1) X (\lambda  X+6 Y)}{3 X+Y}+W+3 X+Y\nonumber\,,\\ 
    \Omega_{m}&=-2 \zeta +\frac{(p-1) X (\lambda  X+6 Y)}{3 X+Y}-V-W-3 X-Y+1\,.
\end{align}
The EoS parameter for DE can be written as.
\begin{align}
   \omega_{DE} =-\frac{(V+3) X+2 Y}{3 X \left(2 \zeta -\frac{(p-1) X (\lambda  X+6 Y)}{3 X+Y}+W+3 X+Y\right)}\nonumber\,.
\end{align}
Next, we have calculated and presented the critical points for this dynamical system in Table \ref{model2criticalpoint}

\begin{figure}[H]
    \centering
    \includegraphics[width=75mm]{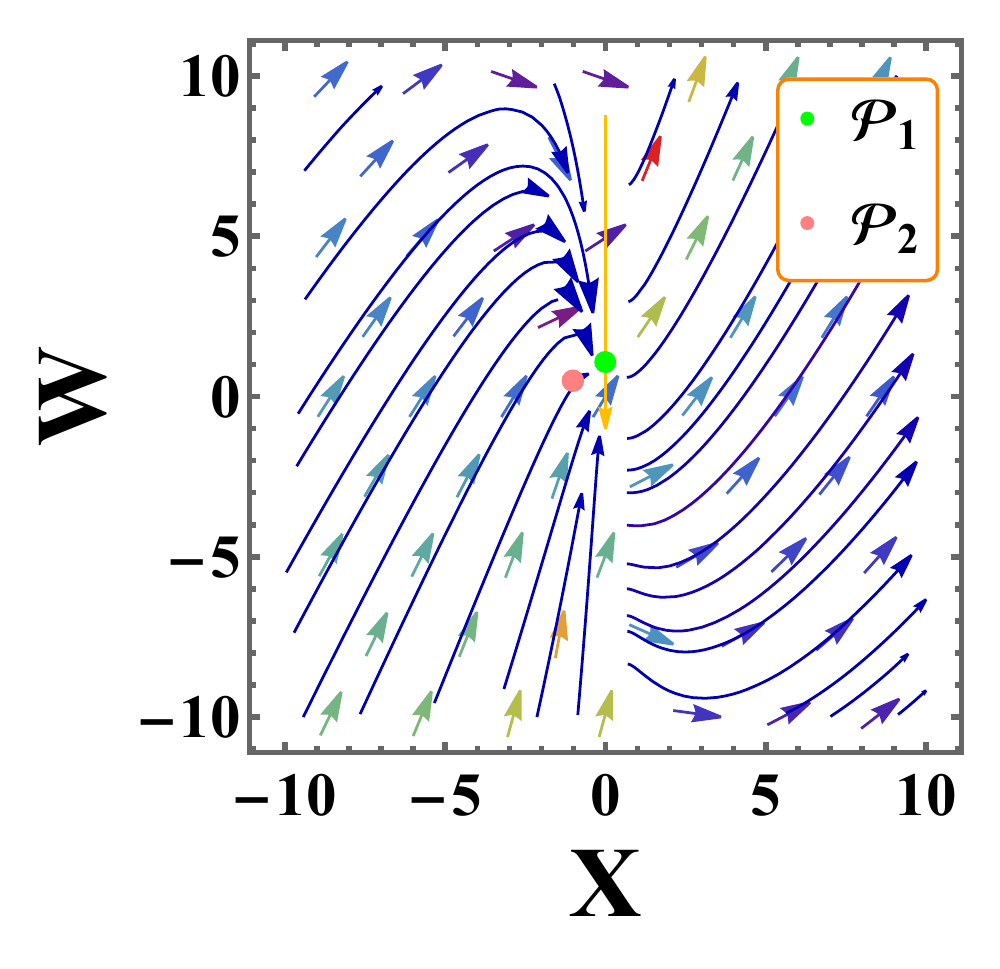}
    \includegraphics[width=75mm]{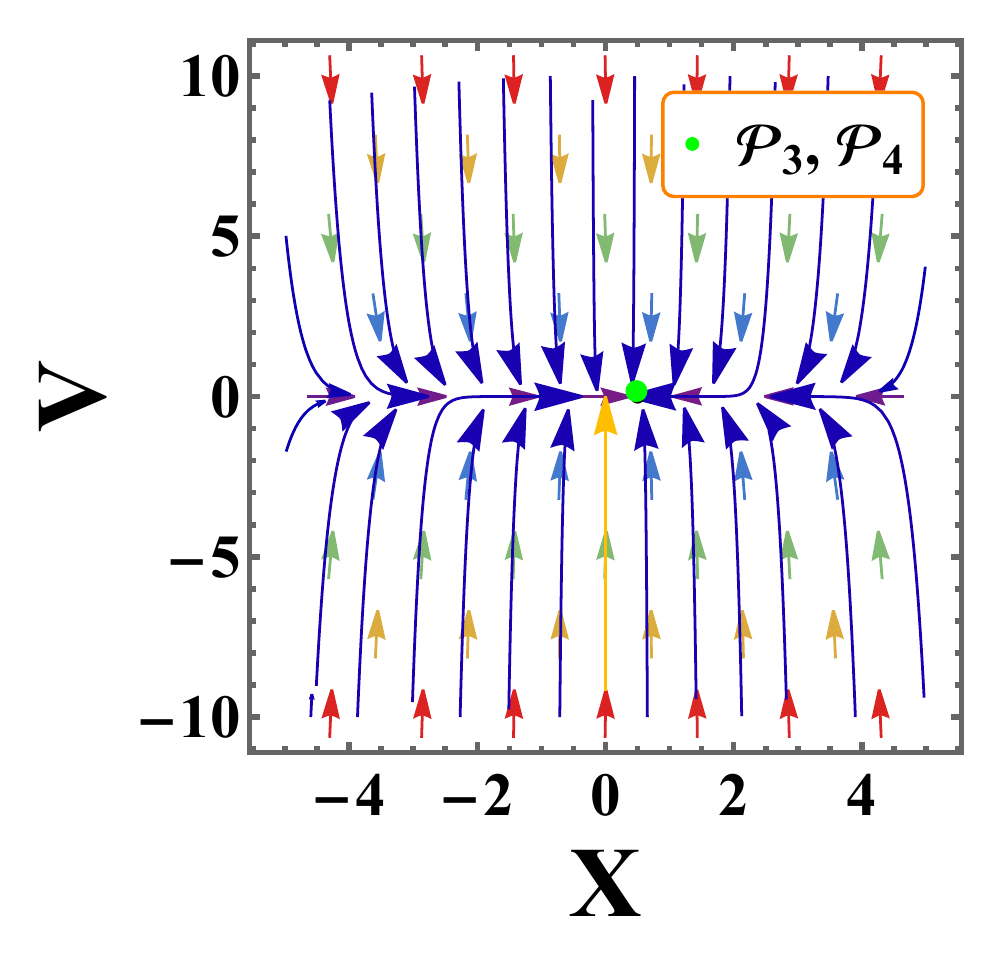}    
    \caption{ $2D$ phase portrait for the dynamical system for $\lambda=0.3,\, \zeta=1.0001,\,p=-1$ ({\bf Model-II}) .} \label{modelII2dphaseportrait}
\end{figure}

\begin{widetext}
\begin{table}[H]
\centering 
\begin{tabular}{|c|c|c|c|c|c|} 
\hline\hline 
 \parbox[c][0.8cm]{3.3cm}{\textbf{Name of Critical Point}} & $\textbf{x}$ & $\textbf{y}$ &$\textbf{v}$ & $\textbf{w}$ & \textbf{Exists for}\\ [0.5ex] 
\hline\hline 
 \parbox[c][1cm]{3.3cm}{\textbf{$\mathcal{P}_1= (x_1,y_1,v_1,w_1)$}} & $x_{1}$ & $-2 x_{1}$  & $v_1$ &$-\zeta -x_1$ & $x_1\neq 0, p=1, \lambda =8.$ \\
\hline
\parbox[c][1cm]{3.3cm}{$\mathcal{P}_2= (x_2,y_2,v_2,w_2)$} & $x_{2}$ & $-\frac{3x_{2}}{2}$  & $0$ & $w_2$ & $ x_2 \ne 0, \zeta =-w_2-\frac{3 x_2}{2}, p=1, \lambda=\frac{9}{2}.$\\
\hline
\parbox[c][1cm]{3.3cm}{$\mathcal{P}_3= (x_3,y_3,v_3,w_3)$} & $x_3$ & $\sqrt{\frac{\lambda}{2}} x_3$  & $0$ &$-\zeta-\frac{1}{2} \left(\sqrt{2\lambda} +6\right) x_3$&$3 x_3+y_3\neq 0, p=1, \lambda=\text{arbitrary}.$ \\
\hline
\parbox[c][1cm]{3.3cm}{$\mathcal{P}_4= (x_4,y_4,v_4,w_4)$} & $x_4$ & $0$  & $0$ &$0$& $ p=\text{arbitrary}, x_4 \ne 0, \lambda=0.$\\
\hline
\end{tabular}
\caption{The critical points (\textbf{Model-II})}
\label{model2criticalpoint}
\end{table}
\end{widetext}
To study the stability of each critical point, the eigenvalues at each critical point of the Jacobian matrix are calculated and presented in Table \ref{modelIIeigenvalues}. Depending upon the sign of the eigenvalues, the stability of the critical point can be concluded. The stability conditions for each critical point, along with the values of $q, \,\omega_{tot}$ and $\omega_{DE}$ are presented in Table \ref{modelIIstabilitycondi} along with the detailed descriptions.
\begin{itemize}
     \item{Critical Point $\mathcal{P}_{1}$:} The value of parameter $\lambda$ is $8$ at $\mathcal{P}_{1}$, hence this critical point represents radiation-dominated era with $\omega_{tot}=\frac{1}{3}$. This critical point is in the standard radiation-dominated era at $V_1=1, \zeta=0$, where $\Omega_{r}=1$ and $\Omega_{m}=\Omega_{DE}=0$ can be observed from Table \ref{modelIIdensityparameter}. The exact solution retraced at this critical point is in the power law $a(t)=t_0 (t)^h$ form with $h=\frac{1}{2}$ which identifies the radiation-dominated era. According to the sign of the eigenvalues presented in Table \ref{modelIIeigenvalues}, this critical point remains as a saddle point and hence is unstable. The same behaviour can be confirmed from the phase space trajectories (Fig. \ref{modelII2dphaseportrait}). The exact cosmological solution at this critical point is shown in Table \ref{modelIIdensityparameter}. 
\end{itemize}
\begin{table}[H]
    \centering 
    \begin{tabular}{|c|c|} 
    \hline\hline 
    \parbox[c][0.3cm]{0.3cm}{\textbf{C.P.}}& \textbf{Eigenvalues } \\ [0.5ex] 
    \hline\hline 
    \parbox[c][0.3cm]{1.3cm}{$\mathcal{P}_1$ } & $\{0,0,4,8\}$ \\
    \hline
    \parbox[c][0.3cm]{1.3cm}{$\mathcal{P}_2$ } & $\{0,-1,3,6\}$  \\
    \hline
   \parbox[c][0.3cm]{1.3cm}{$\mathcal{P}_3$ } &  $\left\{0,-4-\sqrt{2 \lambda},-2 \sqrt{2 \lambda},-\sqrt{2 \lambda}\right\}$   \\
   \hline
   \parbox[c][0.3cm]{1.3cm}{$\mathcal{P}_4$ } &  $\{0,0,0,-4\}$   \\
    \hline
    \end{tabular}
    \caption{Eigenvalues corresponding to each critical point (\textbf{Model-II})}
    \label{modelIIeigenvalues}
\end{table}

\begin{widetext}
\begin{center}
\begin{table}[H]
    \centering 
    \begin{tabular}{|c|c|c|c|c|} 
    \hline\hline 
    \parbox[c][0.8cm]{1.3cm}{ \textbf{C.P.}} & \textbf{Stability Conditions} & $\textbf{q}$ & \textbf{$\omega_{tot}$} & \textbf{$\omega_{DE}$} \\ [0.5ex] 
    \hline\hline 
    \parbox[c][0.8cm]{1.3cm}{\textbf{$\mathcal{P}_1$}} & Unstable & $1$ & $\frac{1}{3}$ & $-\frac{v_1-1}{3 \zeta }$ \\
    \hline
    \parbox[c][0.8cm]{1.3cm}{$\mathcal{P}_2$}  &  Unstable& $\frac{1}{2}$ & $0$& 0\\
    \hline
    \parbox[c][0.8cm]{1.3cm}{$\mathcal{P}_3$}  &  \begin{tabular}{@{}c@{}}     Stable for\\ $x_3\in \mathbb{R}\land \lambda >0$ \end{tabular} & $-1-\sqrt{\frac{\lambda}{2}}$  & $-1-\frac{\sqrt{2 \lambda}}{3} $ & $-\frac{1}{\zeta}+\frac{\sqrt{2 \lambda}}{3 \zeta}$ \\
    \hline
    \parbox[c][0.8cm]{1.3cm}{$\mathcal{P}_4$}  &  $\text{Nonhyperbolic}$ & $-1$  & $-1$ & $-\frac{1}{2 \zeta+w_4+3 x_4}$ \\
    [1ex] 
    \hline 
    \end{tabular}
     \caption{Stability condition, EoS and deceleration parameters (\textbf{Model-II})}
    \label{modelIIstabilitycondi}
\end{table}
\end{center}
\end{widetext}
To identify the phase of the evolution of the Universe and the exact cosmological solutions at the critical points are calculated and presented in Table \ref{modelIIdensityparameter}.
\begin{widetext}
\begin{center}
\begin{table}[H]
    \centering 
    \begin{tabular}{|c|c|c|c|c|c|} 
    \hline\hline 
   \parbox[c][0.6cm]{1cm}{\textbf{C. P.}} & \textbf{Evolution Eqs.} & \textbf{Universe phase} & $\Omega_{m}$& $\Omega_{r}$ & $\Omega_{DE}$\\ [0.5ex] 
    \hline\hline 
   \parbox[c][0.8cm]{1.3cm}{ $\mathcal{P}_1$ }& $\dot{H}=-2 H^{2}$ & $a(t)= t_{0} (2 t+c_{2})^\frac{1}{2}$ & $1-\zeta-v_1$& $v_1$ & $\zeta$\\
    \hline
    \parbox[c][0.8cm]{1.3cm}{$\mathcal{P}_2$} & $\dot{H}=-\frac{3}{2}H^{2}$ & $a(t)= t_{0} (\frac{3}{2}t+c_{2})^\frac{2}{3}$ & $1-\zeta$& $0$ & $\zeta$\\
    \hline
    \parbox[c][0.8cm]{1.3cm}{$\mathcal{P}_3$ }&  $\dot{H}=\sqrt{\frac{\lambda}{2}}H^2$ & $a(t)= t_{0} (-\sqrt{\frac{\lambda}{2}}t+c_{2})^{\sqrt{\frac{2}{\lambda}}}$ & $0$& $0$ & $1$\\
    \hline
    \parbox[c][0.8cm]{1.3cm}{$\mathcal{P}_4$ }&  $\dot{H}=0$ & $a(t)= t_{0} e^{c_2 t}$ & $1-2 \zeta-w_4-3 x_4$& $0$ & $2 \zeta+w_4+3 x_4$\\
    [1ex] 
    \hline 
    \end{tabular}
    \caption{Phase of the Universe, density parameters (\textbf{Model-II}) }
    \label{modelIIdensityparameter}
\end{table}
\end{center}
\end{widetext}

\begin{figure}[H]
    \centering
    \includegraphics[width=75mm]{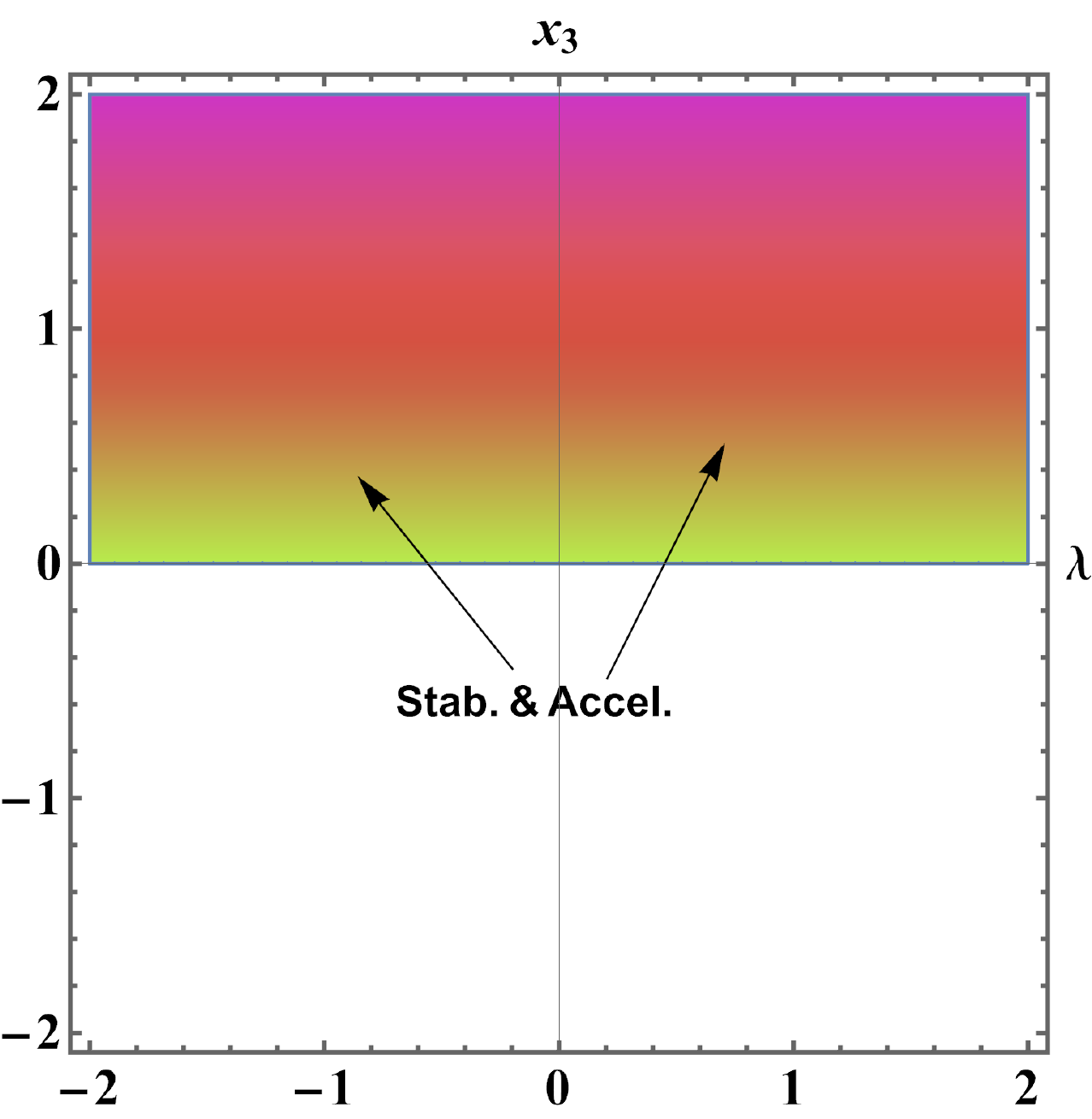}
    \caption{ Stability and the range of parameter $\lambda$ and $x_3$ defining acceleration of the Universe for the critical point $\mathcal{P}_{3}$({\bf Model-II}).} \label{modelIIregionplot}
\end{figure}

\begin{itemize}
\item{Critical Point $\mathcal{P}_{2}$ :} This critical point is representing the CDM-dominated era, where $\omega_{tot}=0$. This can be observed from the exact cosmological solution obtained at this critical point with $a(t)=t_0 (t)^h$, where $h=\frac{2}{3}$  as presented in Table \ref{modelIIdensityparameter}. The value of $\Omega_{m}$ at this critical point is $1$ for $\zeta=0$ and hence represents a standard CDM-dominated era. Since there is an eigenvalue at the Jacobian matrix with a positive sign, this critical point is a saddle point, Ref. Table \ref{modelIIeigenvalues}. From Fig. \ref{modelII2dphaseportrait}, the phase space trajectories at this critical point are moving away from it, hence unstable in behaviour. The parameter $\lambda$ will take the value $\frac{9}{2}$.\\

   \item{Critical Point $\mathcal{P}_{3}$ :} The value of $q$ and $\omega_{tot}$ at this critical point  are $\lambda$ dependent.  Since the value of $\Omega_{DE}=1$, this critical point will represent a standard DE-dominated era and will describe accelerated expansion within the range $x_3\in \mathbb{R}\land \lambda\geq 0$. The stability conditions at this critical point are presented in Table \ref{modelIIstabilitycondi}. For clear visualisation of the parametric range where critical point $\mathcal{P}_3$ is stable and describes accelerating behaviour, we have plotted region plot for parameters $\lambda$ and $x_3$ and is presented in Fig. \ref{modelIIregionplot}. This plot lies in the upper half plane and with the inclusion of $x_3$-axis due to the acceleration range of parameter $\lambda$. We observed that this critical point can describe the de-Sitter solution at $\lambda=0$ and shows stability at the points near $\lambda=0$. The eigenvalues at these critical points are normally hyperbolic and show stability as described in Table \ref{modelIIstabilitycondi}. This critical point is a late time attractor and the same can be visualized from the behavior of phase space trajectories from Fig. \ref{modelII2dphaseportrait}. The exact solution obtained at this critical point is in the power law $a(t)=t_0 (t)^h$ form with $h=\sqrt{\frac{2}{\lambda}}$ which can explains the different epochs of the Universe evolution depending upon value of $\lambda$.

   \item{Critical Points $\mathcal{P}_{4}$ :} This critical point is the de-Sitter solution with $q=\omega_{tot}=-1$. The de-Sitter solution at this critical point is obtained and presented in Table \ref{modelIIdensityparameter}. The standard DE-dominated era can be described by this critical point at $\zeta=\frac{1}{2}, w_4=0, x_4=0$. From the eigenvalues one can see that there are three zeros and one negative eigenvalue, hence linear stability theory will fail to confirm the stability at this point. Therefore, we have moved forward to obtain stability using central manifold theory (CMT). But in this case, while applying CMT we have observed that after co-ordinates shift to the center, this system will not satisfy the CMT condition (after we separated the linear and non-linear parts of the system equations, the nonlinear part is not vanish as same in Model-I critical point $C_3$). We have plotted and analysed the $2-d$ phase portrait at this critical point presented in Fig. \ref{modelII2dphaseportrait}. The phase space trajectories are attracting towards this critical point, hence this critical point is a late time attractor.  
  
\end{itemize}

\begin{figure}[H]
    \centering
    \includegraphics[width=75mm]{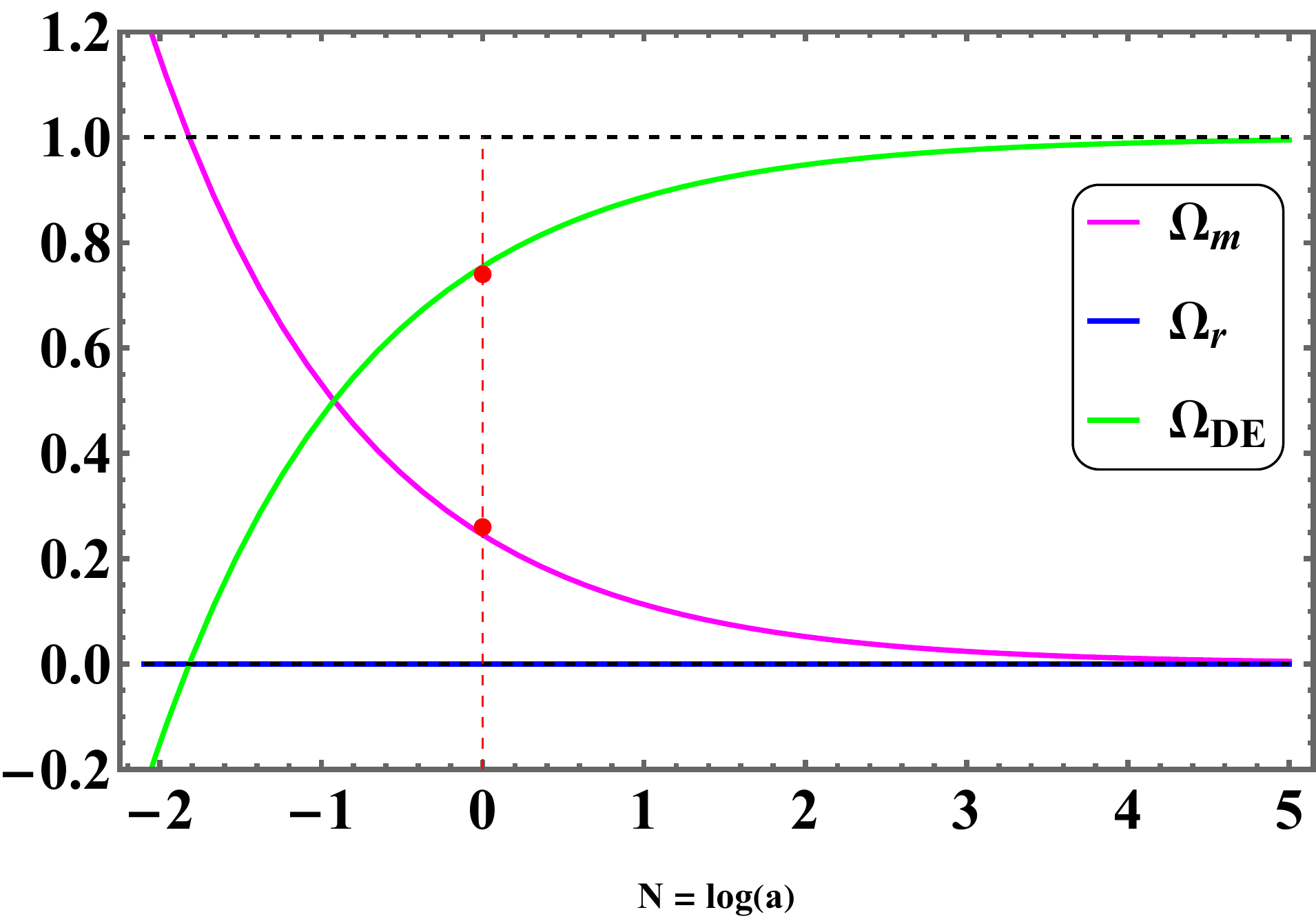}
    \caption{Evolution of the density parameters, with the initial conditions $X=1.2 \times 10^{2.2}, Y=2.2 \times 10^{-3.4}, V=1.02 \times 10^{2.6}, W=4.5 \times10^{-8.1},\, \lambda=0.3,\, \zeta=1.0001,\, p=-1$ (\textbf{Model-II}).} \label{model2evolution}
\end{figure}

\begin{figure}[H]
    \centering
    \includegraphics[width=82mm]{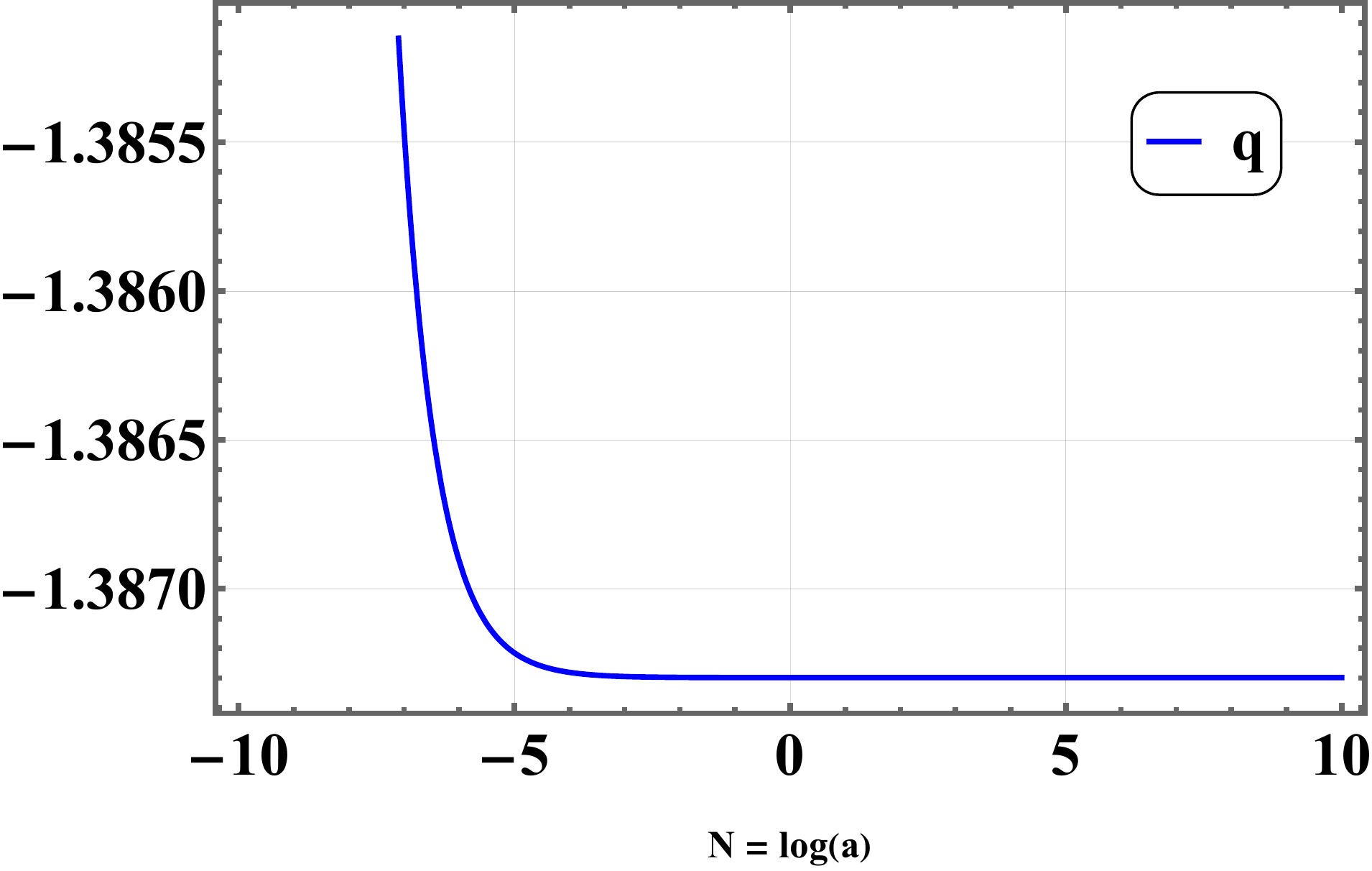}
    \includegraphics[width=75mm]{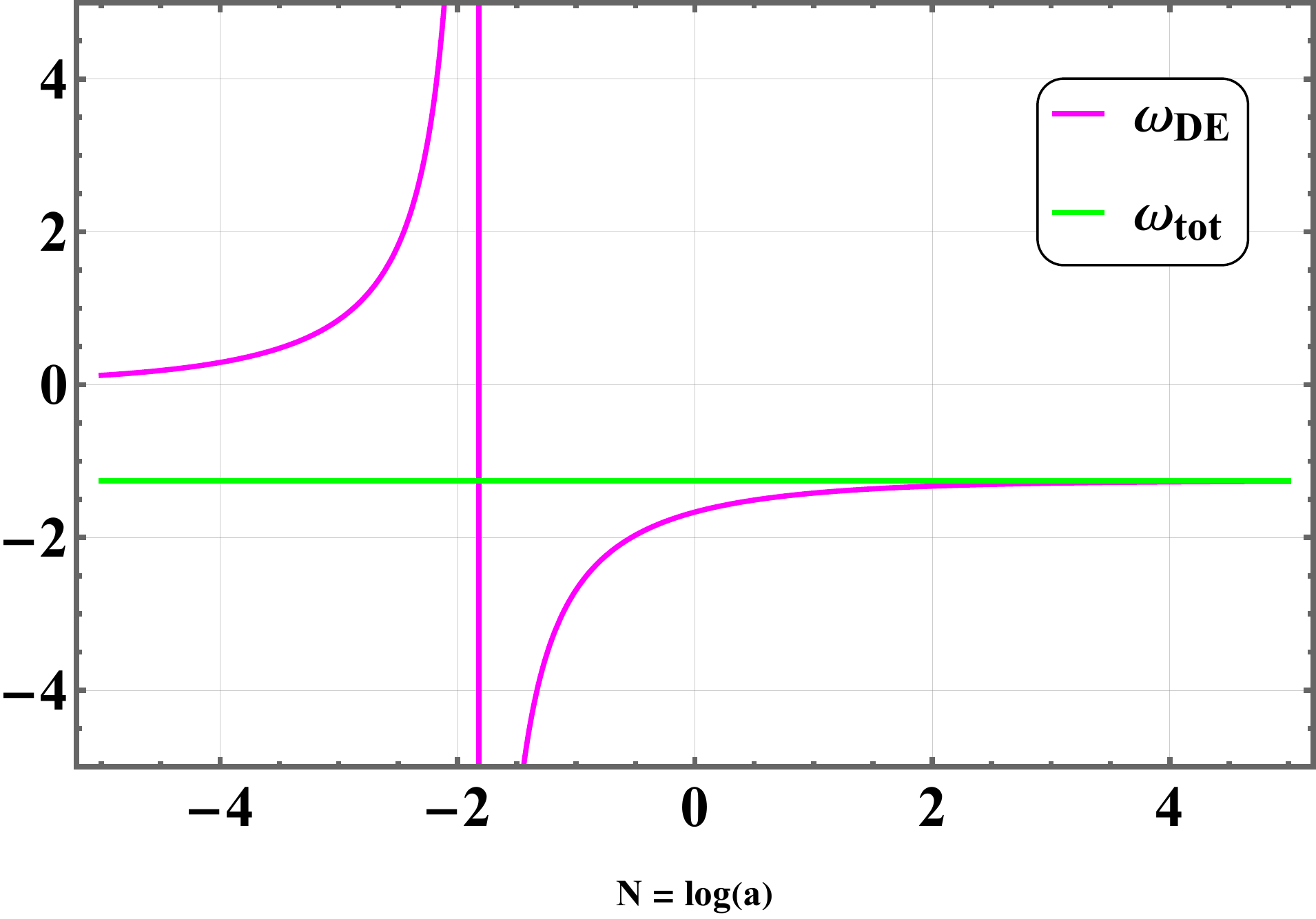}
    \caption{Deceleration $(q)$ and EoS parameters $\omega_{DE}$, $\omega_{tot}$ with the initial conditions $X=1.2 \times 10^{2.2}, Y=2.2 \times 10^{-3.4}, V=1.02 \times 10^{2.6}, W=4.5 \times10^{-8.1},\, \lambda=0.3,\, \zeta=1.0001,\, p=-1$ (\textbf{Model-II}).} \label{model2omegadedecele}
\end{figure}

Graphically we have presented $\Omega_{DE}, \Omega_{m}, \Omega_{r}$ in Fig. \ref{model2evolution} with the initial condition: $X=1.2 \times 10^{2.2}, Y=2.2 \times 10^{-3.4}, V=1.02 \times 10^{2.6}, W=4.5 \times10^{-8.1}$. It has been observed that the value of $\Omega_{m}\approx 0.3$ and $\Omega_{DE} \approx 0.7$ at the present time. The plot for $\Omega_{r}$ vanishes throughout the evolution. The plot for $\Omega_{DE}$ dominates both $\Omega_{r},\,  \Omega_{m}$ at the late phase of cosmic evolution. The plots for $q,\, \omega_{DE},\, \omega_{tot}$ is presented in Fig.\ref{model2omegadedecele}.  The value of $q$ at the present time is $-1.387$ which is approximately the same as the current observation study \cite{Feeney:2018}. The plot of $\omega_{tot}$ at present takes the value $-1.255$ which agrees with \cite{Hinshaw:2013}.


\section{Conclusion}\label{sec:conclusion}
The dynamical system analysis plays an important role in the description of the dynamics of the Universe. The independence of the initial conditions and connecting critical points to the particular evolutionary epochs are important. Hence to analyse the different phases of the Universe's evolution, we have performed the dynamical system analysis for two prominent models in the $f(T, B)$ gravitational theory with the use of the mapping $f(T, B)\rightarrow-T+\Tilde{f}(T, B)$. We present a detailed description of critical points representing all epochs of the evolution of the Universe, the matter-dominated, radiation-dominated, and DE eras. The general dynamical system which is dependent on the form of $\Tilde{f}(T, B)$ is presented in Eq.(\ref{GDS}).  

In subsec. \ref{ModelI}, the form of $\Tilde{f}$ is the linear term of the torsion scalar $T$ along with the logarithmic form of the boundary term $B$. This model is capable of converting a general dynamical system into an autonomous form and is presented in Eq.(\ref{Dynamicalsystemmodel-I}). The critical points along with the existing condition are presented in Table \ref{modelIcriticalpoint}. The stability of each critical point is estimated on the basis of the sign of the eigenvalues of the Jacobian matrix at each critical point. The stability conditions for Model-I are presented in Table \ref{modelIstabilitycondi}. To identify the evolutionary epoch at each critical point, the values of standard density parameters are also calculated and presented in Table \ref{modelIdensityparameter}. From this, it has been observed that for Model-I, the critical point $C_{1}$ representing the radiation-dominated era and the critical point $C_{2}$ representing the matter-dominated era show saddle points and hence are unstable. The same can be verified by analyzing the behavior of phase space trajectories presented in Figs. \ref{modelI2dphaseportrait} and \ref{modelI2dphaseportrait2}. At the critical point $C_{3}$, the value of $\omega_{DE}$ and $\omega_{tot}$ which are dependent on the coordinate $X, Y$, we can study the DE phase of the Universe evolution. For better visibility, the existence and stability region for a critical point $C_3$ is plotted and presented in Fig. \ref{modelIregionplot}. To analyze the behavior of standard density parameters, the plot for $\Omega_{DE}, \Omega_{r}$ and $\Omega_{m}$ in terms of the redshift is presented in Fig. \ref{model1evolution}. From this plot, we can conclude that the density parameter for radiation and matter goes on decreasing and vanishes at a late time, whereas the plot for DE is increasing from an early to a late time of cosmic expansion. The plot for $q$ and $\omega_{DE}$, $\omega_{tot}$  are also presented in Figs. \ref{modelIomegadedecele} and \ref{modelIomegadedecele2} respectively, which show compatibility with the observation study made in \cite{Feeney:2018,DIVALENTINO:2016}.  One of the important findings of the addition of boundary term is that, the  logarithmic form of boundary term is capable to describe a critical point in all the evolutionary phases of the Universe., which might not be possible in the context of $f(T)$ gravity formalism.

Further, we have another form of $\Tilde{f}$ with general index $p$ of the negative of the boundary term $B$. This case enables us to analyse the role of boundary term more clearly in the extended teleparallel gravity and we present the analysis in subsec.-\ref{ModelII}. This form has been widely studied in the literature \cite{Bengochea:2008gz,briffa202}, hence it is interesting to investigate in the higher order gravity. The autonomous dynamical system is presented in Eq.(\ref{Dynamicalsystemmodel-II}) and to analyse the different phases of the Universe's evolution, we obtained the critical points which are presented in Table \ref{model2criticalpoint}. The critical point $\mathcal{P}_3$, the deceleration parameter, and $\omega_{tot}$ show dependence on parameter $\lambda$ whose value contributes to the identification of different phases of the evolution of the Universe. This critical point made a difference in the description of both models to describe the DE-dominated era. The stability conditions are obtained from the eigenvalues and presented in Table \ref{modelIIstabilitycondi} along with the values of $q$, $\omega_{tot}$, $\omega_{DE}$. The stability behaviour has been also confirmed with the behaviour of phase space trajectories presented in Fig. \ref{modelII2dphaseportrait}. In this case, the exact cosmological solutions and the value of $\Omega_{DE}$, $\Omega_{r}$, $\Omega_{m}$ at each critical point have been derived and are presented in Table \ref{modelIIdensityparameter}. From this study, we have concluded that this form is capable to represent and describe all three important phases of the evolution of the Universe.

Both the cosmological models show compatibility in explaining the evolutionary behavior from early to late time. The analysis of the models follow the argument regarding the explanation of the unstable critical points explaining early epochs such as the matter- dominated and radiation-dominated era and the stable behaviour in certain parametric ranges in the de-Sitter phase. The detailed mathematical analysis along with the region of stability is presented with the help of region plots for both models. For a deeper understanding of these critical points and freedom on the model parameter, one may link them with cosmological observations. Also, the models may allow us to differentiate between viable parameter ranges while probing the issues of the early Universe.
\section*{Data Availability Statement:}No data was used for the research described in the article.

\section*{Acknowledgements}
SAK acknowledges the financial support provided by University Grants Commission (UGC) through Senior Research Fellowship (UGC Ref. No.: 191620205335) to carry out the research work. BM acknowledges IUCAA, Pune, India, for hospitality and support during an academic visit where a part of this work has been accomplished. 

\section*{References} 
\bibliographystyle{utphys}
\bibliography{references}
\end{document}